\documentclass[twocolumn,showpacs,amsmath,prd,amssymb,showkeys,superscriptaddress,aps,longbibliography,nofootinbib]{revtex4-2}

\pdfoutput=1

\usepackage{graphicx}
\usepackage{dcolumn}
\usepackage{mathtools}
\usepackage{scalerel}
\usepackage{bm}
\usepackage{upgreek}
\usepackage{mathrsfs}
\usepackage{float}

  \usepackage{array}
 \usepackage{verbatim}
  \usepackage{amstext}
  \usepackage{amsmath}
  \usepackage{amssymb}
  \usepackage{slashed}
  \usepackage[bbgreekl]{mathbbol}
  \usepackage{indentfirst}
  \usepackage{xspace}
 \usepackage[T1]{fontenc}
 \usepackage[latin9]{inputenc}
 \usepackage{color}
 \PassOptionsToPackage{normalem}{ulem}
 \usepackage{ulem}
 \usepackage[unicode=true,pdfusetitle,
  bookmarks=true,bookmarksnumbered=false,bookmarksopen=false,
  breaklinks=false,pdfborder={0 0 1},backref=false,colorlinks=true]
  {hyperref}
 \hypersetup{
  citecolor=blue}

\newcommand{\delcp}         {\delta_{cp}}
\newcommand{\tb}	{\theta_{13}}
\newcommand{\tc}	{\theta_{23}}
\newcommand{\ldm}	{\Delta m_{31}^2}
\newcommand{\axem}	{a^X_{e \mu}}
\newcommand{\axet}	{a^X_{e \tau}}
\newcommand{\axmt}	{a^X_{\mu \tau}}
\newcommand{\cxyem}	{c^{XY}_{e \mu}}
\newcommand{\cxyet}	{c^{XY}_{e \tau}}
\newcommand{\cxymt}{	c^{XY}_{\mu \tau}}

\newcommand{\axab}	{a^X_{\alpha \beta}}
\newcommand{\cxyab}   {c^{XY}_{\alpha \beta}}

\newcommand{\phxab}	{\phi^X_{\alpha \beta}}
\newcommand{\phxyab}   {\phi^{XY}_{\alpha \beta}}

\makeatother

\begin{document}

\title{Investigating Lorentz Invariance Violation Effects on CP Violation and Mass Hierarchy sensitivity at DUNE}

\newcommand{\cusb}{Department of Physics,
Central University of South Bihar, Gaya 824236, India}

\newcommand{\bhu}{Department of Physics, Institute of Science,
Banaras Hindu University,
Varanasi 221005, India.}

\newcommand{\corrskm}{shashankmihsra@cusb.ac.in}
\newcommand{\corrls}{venktesh@cusb.ac.in}

\author{ Saurabh~Shukla} \affiliation{ \cusb }\affiliation{ \bhu }
\author{ Shashank~Mishra} \altaffiliation{ \corrskm }\affiliation{ \cusb }\affiliation{ \bhu }
\author{ Lakhwinder~Singh } \affiliation{ \cusb }
\author{ Venktesh~Singh } \altaffiliation{\corrls}  \affiliation{ \cusb }

\date{\today}

\begin{abstract}
One of the current goals of neutrino experiments is to precisely determine standard unknown oscillation parameters such as the leptonic CP phase and mass hierarchy. Lorentz invariance violation represents a potential physics factor that could influence the experiment's ability to achieve these precise determinations. This study investigates the influence of Lorentz invariance violation (LIV) on oscillation dynamics, particularly through non-isotropic CPT-violating ($a^{X}_{e\mu}$, $a^{X}_{e\tau}$, $a^{X}_{\mu\tau}$) and CPT-conserving ($c^{XY}_{e\mu}$, $c^{XY}_{e \tau}$, $c^{XY}_{\mu \tau}$) parameters within the Deep Underground Neutrino Experiment (DUNE).
We analyze the impact of these parameters on the mass hierarchy (MH) and Dirac CP phase sensitivity measurements. Our findings indicate that while MH sensitivity remains relatively unaffected, only the presence of $c^{XY}_{\mu \tau}$ significantly deteriorates MH sensitivity, albeit remaining above the $5 \sigma$ threshold. Additionally, we observe a substantial compromise in CP sensitivity due to the $c^{XY}_{e \mu}$ and $c^{XY}_{e \tau}$ parameters.
 \end{abstract}

\maketitle

\section{Introduction}
The Standard Model (SM) is a experimentally tested theory that describes elementary particles and their interactions, with the exception of gravity. However, there are significant gaps remain that require attention. One such gap is phenomenon of neutrino oscillation, where inter-conversion among three flavors suggests the presence of neutrino mass. However, the SM posits massless neutrinos. Introducing neutrino mass necessitates the incorporation of new physics beyond the standard model (BSM). 

The parameters constituting the three-neutrino oscillation framework encompass the three mixing angles $\theta_{12}, \theta_{13}$, and $\theta_{23}$, the two mass squared differences, denoted as $\Delta m^{2}_{31} = m^{2}_{3} -m^{2}_{1} $ and $\Delta m^{2}_{21} =m^{2}_{2} - m^{2}_{1}$ in addition to the Dirac CP phase $\delta_{cp}$. Presently, The parameters $\theta_{12}$, $\Delta m^2_{21}$ (including its sign), and $\theta_{13}$ have been precisely measured, while we also possess a solid understanding of the magnitude of $\Delta m^2_{31}$~\cite{Workman:2022ynf}.
Current and next-generation proposed neutrino oscillation experiments primarily aim to precisely determine critical parameters, including $\theta_{23}$, the leptonic CP phase $\delta_{cp}$, and the atmospheric mass squared difference $\Delta m^{2}_{31}$~\cite{Esteban:2020cvm}. 

The leptonic sector may have potential CP Violation (CPV) if $\delcp$ deviates from $0$ or $\pi$. This could be the key for unraveling the Universe's matter-antimatter asymmetry puzzle~\cite{FUKUGITA198645,RevModPhys.84.515}.
 Another crucial inquiry pertains to determining the neutrino mass hierarchy, whether it follows a normal (NH, $\Delta m^2_{31} > 0$) or inverted (IH, $\Delta m^2_{31} < 0$) hierarchy. Such investigations not only shed light on plausible models for neutrino mass generation but also helps in discerning the nature of neutrino (Dirac or Majorana)~\cite{HAXTON1984409}. 
Several upcoming next generation experiments are in position to improve the accuracy of critical parameters. These efforts also hold promise for the exploration of various scenarios of BSM. Investigating potential deviations from standard behavior in neutrinos is a well-established approach to delve deeper into new physics. These non-standard effects could influence how accurately experiments measure neutrino oscillation parameters, thereby affecting the overall performance of neutrino experiments ~\cite{Fiza:2022xfw,KumarAgarwalla:2019gdj,Masud:2016nuj,Masud:2016bvp,Pan:2023qln,Raikwal:2023lzk}.

Lorentz symmetry is crucial in the construction of quantum field theory within the Standard Model (SM), which uses gauge theory to describe the interactions among fundamental particles.
When the fundamental symmetries of space-time are disrupted, it leads to what is known as a Lorentz invariance violation (LIV). 
 In higher-dimensional theories associated to the Planck scale, Lorentz invariance violation can emerge spontaneously ~\cite{Kostelecky:1988zi,Kostelecky:1991ak,Colladay:1998fq,Bluhm:2005uj}. Neutrino oscillation experiments offer a critical platform for probing Lorentz invariance violation ~\cite{Diaz:2009qk,Kostelecky:2003xn,Kostelecky:2004hg}. This violation can manifest isotropically or non-isotropically, with the latter suggesting directional asymmetry in space-time disruption. 

Non-isotropic LIV, in conjunction with the sidereal effect, posits directional variations in space-time symmetries relative to celestial structures. The Earth's motion through space thus presents a dynamic laboratory, where experiment's orientations relative to celestial backgrounds may influence observable neutrino behaviors, including propagation speed and flavor oscillations.
Numerous studies have been conducted across various experiments to explore and understand the Lorentz invariance violation for the isotropic ~\cite{MINOS:2008fnv,Super-Kamiokande:2014exs}  and nonisotropic ~\cite{LSND:2005oop,MINOS:2010kat,MINOS:2012ozn,MiniBooNE:2011pix,IceCube:2010fyu,DoubleChooz:2012eiq,T2K:2017ega} scenarios. Several phenomenological studies have been performed to study the implication of isotropic LIV in various long baseline experiments ~\cite{Pan:2023qln,Raikwal:2023lzk}.

In this study, we have investigated the influence of non-diagonal non-isotropic LIV parameters on the potential of the Long-Baseline (LBL) experiment through the sidereal effect. We focus the proposed next-generation LBL experiment, Deep Underground Neutirno Experiment(DUNE)~\cite{DUNE:2020lwj}. The baseline chosen for DUNE has been specifically optimized to enhance the sensitivity of the experiment to the CP Violation and  and is also ideally suited for addressing the question of neutrino mass ordering in the SM scenario~\cite{DUNE:2020jqi}. We examine the  mass hierarchy and CP-violation sensitivity in the presence of LIV. A detailed CP-precision study has been carried out to understand how LIV influences the constraints on the $\delta_{CP}$ phase at DUNE. We highlight the salient characteristics of the sensitivity analysis with respect to different LIV parameters in our discussion. It is important to understand that the inclusion of
new-physics scenarios may cause deviation in the precise determination of mixing parameters in LBL experiments.

The manuscript is structured as follows: Section ~\ref{sec::formulation} presents a brief overview of the theoretical framework regarding Lorentz invariance violation (LIV). Section ~\ref{sec::Exp} details the experimental and simulation methodologies. Section ~\ref{sec::Prob} discusses the probabilistic aspects in the context of LIV parameters. Section ~\ref{sec::sensitivity} outlines our methodology for conducting sensitivity analysis. The main findings are elaborated in Section ~\ref{sec::results}, where we provide a qualitative discussion of our results. Finally, Section ~\ref{sec::summary} provides a summary of our work.

\section{Formalism}
\label{sec::formulation}

In the Standard model extension the Lorentz invariance violating Lagrangian density for the neutrinos(antineutrinos) can be described as~\cite{PhysRevD.69.016005,PhysRevD.85.096005}
\begin{equation}
    \mathcal{L}=\frac{1}{2}\bar{\psi}(i\partial-M-\hat{\mathcal{Q}})\psi + h.c.
\label{lagliv}
\end{equation}

Here, The lagrangian consists of three major terms: first terms is a kinetic term and second term is a mass term and the third term represended by $\hat{\mathcal{Q}}$ is lorentz violating operator for neutrino (antineutrinos) fields  denoted by fermionic spinor  $\psi$ ($\bar{\psi}$). This LIV opetaor, incapulalting the Lorentz invariant violating part can be represented as follow when renormalizable dirac coupling is taken into account~\cite{PhysRevD.85.096005}: 

  \begin{equation}
    \begin{aligned}[b]
        & \mathcal{L}_{\rm LIV} =\frac{-1}{2}\left[a^{\mu}_{\alpha\beta} \bar{\psi_{\alpha}}\gamma_{\mu}\psi_{\beta}+ b^{\mu}_{\alpha\beta} \bar{\psi_{\alpha}}\gamma_{5}\gamma_{\mu}\psi_{\beta}\right]\\
        & \frac{-1}{2}\left[-ic^{\mu\nu}_{\alpha\beta} \bar{\psi_{\alpha}}\gamma_{\mu}\partial_{\nu}\psi_{\beta} - id^{\mu\nu}_{\alpha\beta} \bar{\psi_{\alpha}}\gamma_{5}\gamma_{\mu}\partial_{\nu}\psi_{\beta} \right] + h.c
    \end{aligned}
  \end{equation}
  
Newly defined coefficients, represented as follows:

\begin{equation}
(a_{L})^{\mu}_{\alpha\beta}=(a + b )^{\mu}_{\alpha\beta}, \hfill    (c_{L})^{\mu\nu}_{\alpha\beta}=(c  + d )^{\mu\nu}_{\alpha\beta},
\end{equation}

which are constant Hermitian matrices in the flavor space and have the ability to alter the standard vacuum Hamiltonian, govern the observable effects on left-handed neutrinos. The LIV lagrangian comprises two terms: the CPT-even LIV term  $(a_{L})^{\mu}_{\alpha\beta}$ and the CPT-odd LIV term $(c_{L})^{\mu\nu}_{\alpha\beta}$.

Explicitly, one can write the Lorentz violating contribution to the total oscillation Hamiltonian as,

\begin{equation}
    H=UMU^{\dagger}+V_{m}+H_{\rm LIV},
\end{equation}

The Pontecorvo-Maki-Nakagawa-Sakata (PMNS) mixing matrix, denoted as $U$ \cite{Kopp:2007ne}, incorporates three mixing angles $\theta_{12}$, $\theta_{13}$, and $\theta_{23}$, along with one CP-violating phase $\delta_{\rm cp}$. The matrix $M$ represents the neutrino mass matrix and is expressed in terms of the mass-squared differences $\Delta m^2_{21}$ and $\Delta m^2_{31}$.

\begin{equation}
    M=\frac{1}{2E}\begin{pmatrix}
0 & 0 & 0\\
0 & \Delta m^{2}_{21} & 0\\
0 & 0 & \Delta m^{2}_{31}
\end{pmatrix}.
\end{equation}

The matter potential  $V_{m}$ is given by
 
\begin{equation}
    V_{m}=\pm\sqrt{2}G_{F}N_{e}\begin{pmatrix}
1 & 0 & 0\\
0 & 0 & 0\\
0 & 0 & 0
\end{pmatrix}.
\end{equation}

where $G_{F}$ stands for Fermi constant and $N_{e}$ refers to the electron density in the matter. The $+$ sign in $V_{m}$ is for neutrinos and $-$ sign is for antineutrinos. The LIV-Hamiltonian ($H_{\rm LIV}$) for neutrino-neutrino \footnote{In the case of antineutrino-antineutrino mixing, $(a_{L})^{\mu}$ changes to $- ((a_{L})^{\mu})^{*}$, $(c_{L})^{\mu\nu}$ changes to $ ((c_{L})^{\mu\nu})^{*}$}
 mixing is given by ~\cite{Kostelecky:2003xn}
 
\begin{equation}
  (\mathcal{H}_{LIV})_{\alpha\beta}  = \frac{1}{E}[(a_{L})^{\mu} p_{\mu}  - (c_{L})^{\mu\nu} p_{\mu} p_{\nu}]_{\alpha \beta}.
  \label{hLIV}
\end{equation}

Where $(a_{L})^{\mu}$ and $(c_{L})^{\mu\nu}$ are 3$\times$3 complex matrices represent LIV coefficients with mass dimension 1 and 0, respectively \footnote{Later in the paper, we have omiited the subscript 'L' from the parameters.}. 
Here indices $\alpha$, $\beta$ = e, $\mu$, $\tau$ refers to the flavors of neutrinos.
Search of Non-isotropic LIV predominantly relies on the change in the direction of the beam, which is directly achieved by rotation of the earth. To get the absolute measure of rotation, reference is typically made to distant stars, and such rotation of earth about its axis using distant star as point of measure, is called sidereal rotation. Equation~\ref{hLIV} can express the directional dependence in terms of this sidereal rotation. For earth based searches, the standard inertial frame commonly used is the sun-centered celestial-equatorial frame (SCCEF), described with coordinates (X, Y, Z, T) ~\cite{PhysRevD.66.056005}. In experiments conducted on Earth, both the source and detector rotate at an angular frequency approximately equal to 2$\pi$/(23 h 56 min).

The sidereal time dependence of the effective Hamiltonian (Eq. 3 of ~\cite{Kostelecky:2004hg}) in this scenario can be explicitly represented by the following equation~\cite{PhysRevD.109.075042}:
\setlength\abovedisplayskip{5pt}
\begin{widetext}
\begin{equation}
  \begin{split}
       (\mathcal{H}_{LIV})_{\alpha\beta} = (C)_{\alpha\beta} + R [ a_{\alpha\beta}^{X} - 2 E (c^{TX})_{\alpha\beta} + 2 E N_{z} (c^{XZ})_{\alpha\beta}] sin(\omega_{\oplus} T - \Phi_{orientation})~- \\
   R [ a_{\alpha\beta}^{Y} - 2 E (c^{TY})_{\alpha\beta} + 2 E N_{z} (c^{YZ})_{\alpha\beta}] cos(\omega_{\oplus} T - \Phi_{orientation})~+ \\
   R^{2} [ E \frac{1}{2}((c^{XX})_{\alpha\beta} - (c^{YY})_{\alpha\beta} ) ] cos(2 (\omega_{\oplus} T - \Phi_{orientation}))~+ \\
   R^{2} [ E (c^{XY})_{\alpha\beta}] sin(2 (\omega_{\oplus} T - \Phi_{orientation})),
  \end{split}
  \label{hLIVexpanded}
\end{equation}
\end{widetext}

where $T$ represents the sidereal time, indicating Earth's rotation relative to a sidereal star in the sun-centered frame. $\Phi_{orientation}$ and $R$ can be described using directional factors $N^{X}$, $N^{Y}$, and $N^{Z}$ as follows: 

$$  \Phi_{orientation} = \tan^{-1}(N^{X}/N^{Y}),\\$$
$$  R = \sqrt{ N_{X}^2 + N_{Y}^2 }. $$

The directional factors ($N^{X}$, $N^{Y}$, $N^{Z}$) correspond to the orientation of the beam and the position of the detector. They are defined in terms of the following angles: the Zenith angle ($\theta$), which denotes the angle between the beam and the vertical upward direction; the bearing angle ($\phi$), indicating the angle between the beam and the south direction measured towards the east; and the colatitude ($\chi$) of the detector, which is the complement of the latitude of the detector location on Earth, measured from the north pole~\cite{Kostelecky:2004hg}. 

\begin{equation}
  \begin{split}
  N^{X} &= \cos\chi\sin\theta\cos\phi + \sin\chi\cos\theta,\\
  N^{Y} &= \sin\theta\sin\phi,\\
  N^{Z} &= -\sin\chi\sin\theta\cos\phi + \cos\chi\cos\theta,.\\
  \end{split}
  \label{orientationbeam}
\end{equation}
 The LIV coefficients $(a)_{\alpha \beta}^{\mu}$ are solely governed by the baseline, while  coefficients $(c)_{\alpha\beta}^{\mu\nu}$ are subject to control from both the baseline length and the energy of the neutrinos.
In this work, we examine the effects of the non diagonal CPT violating LIV parameters $a^{X}_{e\mu}, a^{X}_{e \tau},  a^{X}_{\mu \tau}$ and CPT conserving LIV parameters $c^{XY}_{e\mu}, c^{XY}_{e \tau} , c^{XY}_{\mu \tau}$ on the oscillation probabilities. In this paper, we study the impact of an individual parameter; hence, only one parameter is considered non-zero at a time. Now in this paper, we use $\axab$ and $\cxyab$ as the magnitude of the LIV parameter while $\phxab, \phxyab$ as the LIV phase of the parameter.
Current data limits on the LIV parameters of the neutrino sector can be found at~\cite{RevModPhys.83.11}
\section{Experimental Setup and Simulation}
\label{sec::Exp}

\begin{table}[H]
  \caption{The standard oscillation parameters are  used in this work.(ref)}
  \label{table2OScPar}
     \setlength{\tabcolsep}{8pt}
 \renewcommand{\arraystretch}{1.3}
  \begin{tabular}{|c|c|c|}
    \hline 
    Parameter            & True Value      &  Marginalization Range\\ \hline \hline
    $\theta_{12}$         & $33.82^{\circ}$ &  -- \\ \hline
    $ \theta_{13}$        & $8.5^{\circ}$   &  $(8.43,8.65)$\\ \hline
    $ \theta_{23}$        & $49.0^{\circ}$  &  $(41.0^{\circ},52.0^{\circ}) $\\ \hline
    $\delta_{cp}$         &$ 195.0^{\circ}$  &  $[-\pi,\pi] $ \\ \hline
    $\triangle m^{2}_{21}$    & $7.39 \times 10^{-5} eV^{2}$  & -- \\ \hline 
    $\triangle m^{2}_{31}$    & $2.45  \times 10^{-3} eV^{2}$  & $(2.36 \times 10^{-3}, 2.64 \times 10^{-3}) $\\ \hline
    $|a^X_{\alpha \beta}|$    & $5.0 \times 10^{-23}  GeV $  & --\\ \hline
      $\phi^X_{\alpha \beta}$   &  --  & $[-\pi,+\pi]$\\ \hline
      $c^{XY}_{\alpha \beta}$  &  $ 5.0 \times 10^{-23} $  & --\\ \hline
     $\phi^{XY}_{\alpha \beta}$ &   --  & $[-\pi,+\pi]$\\ \hline
     \hline
     \end{tabular}
\end{table}

The Deep Underground Neutrino Experiment (DUNE) involves two detectors exposed to a megawatt-power muon neutrino beam produced at Fermilab.
 DUNE consists of two detectors: a near detector located close to the muon neutrino beam source, and a remote detector consisting of four 10-kiloton liquid argon time projection chambers (TPCs) located 1300 km away at the Sanford Underground Research Facility in South Dakota.
The experiment aims to achieve several scientific goals, including measuring leptonic CP violation, determining the order of neutrino masses, and precisely determining neutrino mixing parameters. To simulate DUNE, we utilize the GLoBES software package with the latest DUNE configuration file provided by the collaboration \cite{DUNE:2021cuw}, running simulations for 5 years in neutrino mode and 5 years in antineutrino mode to model the anticipated experimental results.
The analysis includes both disappearance and appearance channels. Modifications have been made to the snu.c plugin to incorporate the sidereal effect, as described in reference \cite{Kostelecky:2004hg}. 
\section{Probabilities in the presence of LIV}
\label{sec::Prob}
The oscillation probabilities induced by Lorentz invariance violation (LIV) in both the appearance ($\mu \rightarrow e$) and disappearance ($\mu \rightarrow \mu$) channels can be expressed up to the leading order, analogous to the formulation detailed in the ref ~\cite{Liao:2016hsa,Chaves:2018sih,Yasuda:2007jp}
\begin{widetext}
  \begin{equation}
  \begin{split}
     P_{\mu e}^{\rm LIV}  \simeq x^2 f^2 +2 x y fg \cos (\Delta+\delta_{CP})
    + y^2 g^2 + 4 r_A |{h}^{\rm{LIV}}_{e \mu}| 
    \big\{ xf \big [ f s_{23}^2 \cos (\phi^{\rm{LIV}}_{e \mu}+\delta_{CP})
      +g c_{23}^2 \cos( \Delta +\delta_{CP}+\phi^{\rm{LIV}}_{e \mu})\big] \\
     + yg \big[ g c_{23}^2 \cos \phi^{\rm{LIV}}_{e \mu} 
      + f s_{23}^2\cos (\Delta-\phi^{\rm{LIV}}_{e \mu})\big ]\big \}
    + 4 r_A |{h}^{\rm{LIV}}_{e\tau}| s_{23} c_{23} \big\{xf \big[ f \cos (\phi^{\rm{LIV}}_{e \tau}+\delta_{CP})
      - g\cos (\Delta +\delta_{CP}+\phi^{\rm{LIV}}_{e \tau}) \big] \\
       - yg[ g \cos \phi^{\rm{LIV}}_{e \tau}
      -f \cos(\Delta-\phi^{\rm{LIV}}_{e\tau})\big]\big\}+ 4 r_A^2 g^2 c_{23}^2 | c_{23} |{h}^{\rm{LIV}}_{e \mu}|
    -s_{23} |{h}^{\rm{LIV}}_{e\tau}||^2 + 4 r_A^2 f^2 s_{23}^2  | s_{23} |{h}^{\rm{LIV}}_{e \mu}|
    +c_{23} |{h}^{\rm{LIV}}_{e\tau}||^2 \\
     + 8 r_A^2 f g s_{23}c_{23} \big\{ c_{23} \cos \Delta \big[s_{23}(|{h}^{\rm{LIV}}_{e \mu}|^2 -|{h}^{\rm{LIV}}_{e \tau}|^2)
      + 2 c_{23} |{h}^{\rm{LIV}}_{e \mu}| |{h}^{\rm{LIV}}_{e \tau}|\cos(\phi^{\rm{LIV}}_{e \mu}-\phi^{\rm{LIV}}_{e \tau}) \big] \\
     - |{h}^{\rm{LIV}}_{e \mu}|| {h}^{\rm{LIV}}_{e \tau}| \cos (\Delta -\phi^{\rm{LIV}}_{e \mu}
    +\phi^{\rm{LIV}}_{e \tau})\big \} 
    +{\cal O}(s_{13}^2 a, s_{13}a^2, a^3),
    \label{pmue}
\end{split}
\end{equation}
\begin{equation}
  \begin{split}
    P_{\mu\mu}^{\rm LIV} & \simeq 1- \sin^2 2 \theta_{23}\sin ^2 \Delta  - |{h}^{\rm{LIV}}_{\mu\tau}|
    \cos \phi^{\rm{LIV}}_{{\mu\tau}} \sin 2 \theta_{23}
    \Big[ (2r_A\Delta )\sin^2 2\theta_{23}\sin 2\Delta + 4 \cos^2 2 \theta_{23}r_A\sin ^2\Delta\Big]\\
    & + (|{h}^{\rm{LIV}}_{\mu\mu}| - |{h}^{\rm{LIV}}_{\tau\tau}|)\sin^2  2 \theta_{23} \cos 2 \theta_{23}
    \Big[(r_A\Delta) \sin 2\Delta -2r_A \sin ^2 \Delta  \Big],
    \label{pmumu}
  \end{split}
\end{equation}
where
\begin{equation}
  \begin{split}
    s_{ij}=\sin\theta_{ij},~~c_{ij}=\cos\theta_{ij},~~ 
    x=2s_{13}s_{23},~~ y=2rs_{12}c_{12}c_{23},~~  r=|\Delta m^2_{21}/\Delta m^2_{31}|,~~
     \Delta = \frac{\Delta m^2_{31} L}{4E},~~   \\
     V_{CC}=\sqrt 2 G_F N_e,~~ 
     r_A=\frac{2E}{{\Delta m}^2_{31}},~~
    f=\frac{\sin\big[\Delta (1-r_A(V_{CC}+{h}^{\rm{LIV}}_{ee}))  \big]}{1-r_A(V_{CC}+{h}^{\rm{LIV}}_{ee})},~~
    g=\frac{\sin\big[\Delta r_A(V_{CC}+{h}^{\rm{LIV}}_{ee})  \big]}{r_A(V_{CC}+{h}^{\rm{LIV}}_{ee})}.\\
    \hspace{0.5 true cm}\label{os-po}
  \end{split}
\end{equation}
\end{widetext}
The antineutrino probability can be obtained by replacing $V_{CC} \rightarrow -V_{CC}$, $\delcp \rightarrow -\delcp$ and ${h}^{\mathrm{LIV}}_{\alpha \beta}$ to ${{h}^{\mathrm{LIV}}_{\alpha \beta}}^\ast$.
Here \( |{h}^{\mathrm{LIV}}_{\alpha \beta}| \) and \( |{\phi}^{\mathrm{LIV}}_{\alpha \beta}| \) refer to the magnitude and phase of the element of the LIV Hamiltonian matrix.
Based on the above formula, it's crucial to emphasize that when LIV parameters are present, the probability of neutrino appearance relies on $e\mu$ and $e\tau$ type parameters ($\axem, \axet, \cxyem, \cxyet$). This probability is also influenced by LIV phases 
$\phi_{e \mu}$ and $\phi_{e \tau}$, particularly in conjunction with 
the Dirac CP phase $\delta_{cp}$. Conversely, in the disappearance channel, the probability depends on $\mu \tau$ type parameters ($\axmt,\cxymt$). The disappearance probability does not depend on the Dirac CP-phase at the leading order. Therefore, in this case, there is no link between Dirac CP-phase $\delta_{cp}$ and LIV phase $\phi_{\mu \tau}$.
\subsection{Bi-Probability Plots and LIV}
\begin{widetext}

 \begin{figure}[H]
 \centering
       \includegraphics[height= 0.60\textwidth,width=0.95\textwidth]{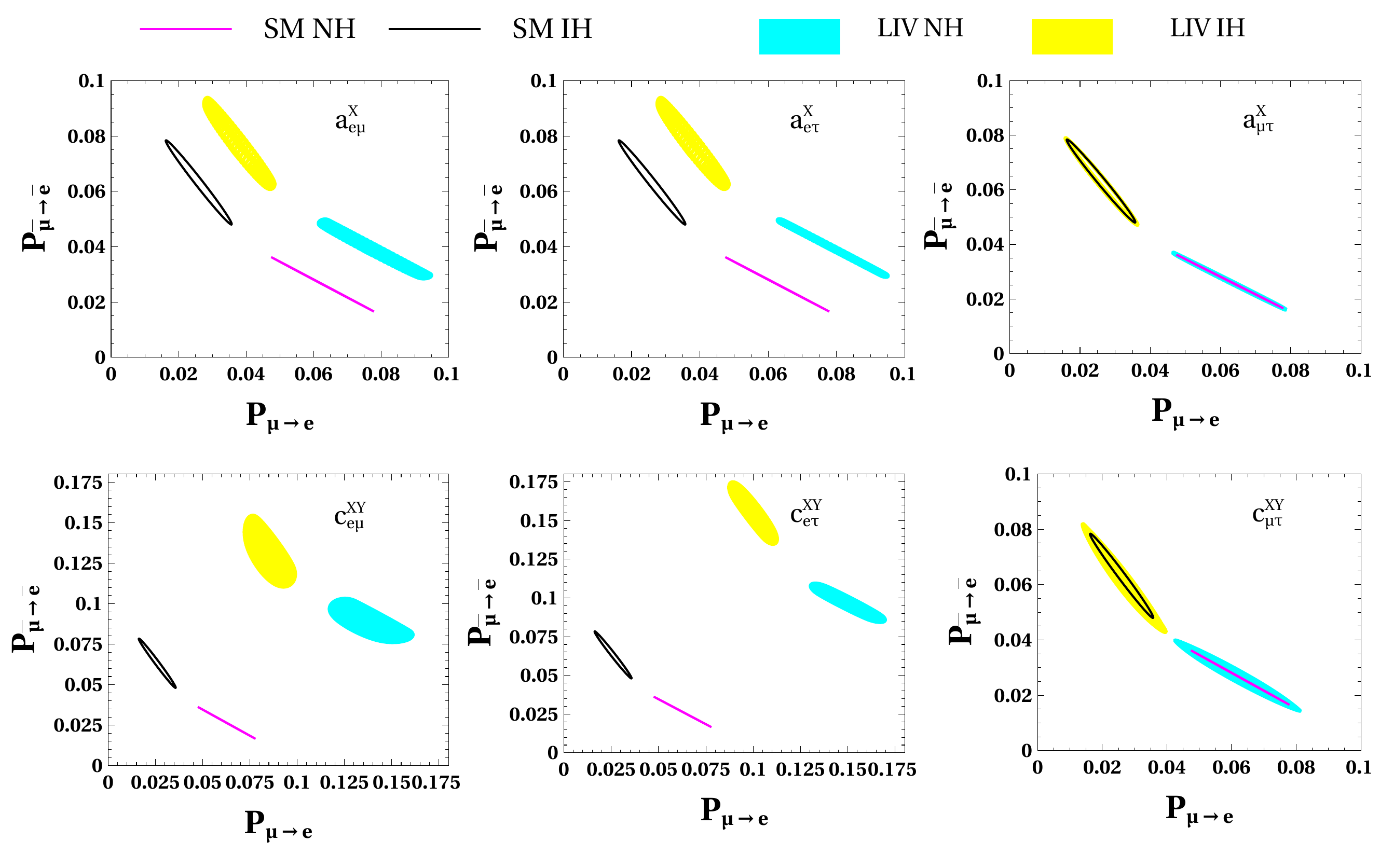}
          \caption{ Bi-probability plots due to variation of $\delta_{cp}$ at $E = 2.5$ GeV, comparing the Standard Model (SM, represented by solid lines) and SM + LIV (depicted by color shaded regions) for DUNE. The top panel, from left to right, corresponds to parameters $a^{X}_{e\mu}$, $a^{X}_{e\tau}$ and $\axmt$, while the bottom panel, also from left to right, corresponds to parameters $c^{XY}_{e\mu}$, $c^{XY}_{e\tau}$, and $c^{XY}_{\mu\tau}$. In the LIV case, parameters $a^{X}_{\alpha\beta} = 1 \times 10^{-23}$ GeV and $c^{XY}_{\alpha\beta} = 1 \times 10^{-23}$ GeV are assumed.}
   \label{biprobb}
    \end{figure} 
    
    	 \end{widetext}

Bi-probability plots of $P(\nu_\mu \rightarrow \nu_e)$ and $P(\bar{\nu}_\mu \rightarrow \bar{\nu}_e)$ are valuable tools for examining the effects of the CP-violating phase and neutrino mass hierarchy (normal or inverted hierarchy) on oscillation probabilities.
Figure~\ref{biprobb} illustrates bi-probability plots, where axes represent probabilities for neutrino and antineutrino cases at a fixed energy of E=2.5 GeV\footnote{The first vacuum oscillation maxima for the appearance channel occurs at 2.5 GeV for standard conditions in the case of DUNE}.
In the context of the Standard Model (SM), ellipses arise from the modulation of the Dirac CP-phase.
When additional Lorentz Invariance Violation (LIV) is included, the resulting SM+LIV blobs are represented as a scatter plot, reflecting simultaneous variations in both the standard Dirac CP-phase and the LIV phase.
Bi-probability plots are constructed under both NH and IH scenarios, assuming equal strengths for all parameters $\axab$(in GeV) and $\cxyab$ at $5.0 \times 10^{-23}$. For the SM case in DUNE, there is no mass hierarchy degeneracy, as the ellipses do not overlap. When $\axab$ parameters are introduced, the probabilities change without overlapping between NH and IH plots. However, with $\cxyem, \cxyet$ type parameters, the probabilities change universally, and for the $\cxymt$ parameter, MH curves approach each other compared to the SM scenario\footnote{If current limit of the $\cxymt$ is taken then there would be a larger degeneracy}. Impact of the $\cxyab$ parameters is more than $\axab$ parameters due to their direct dependency on energy. Conjuction between Dirac CP-phase $\delta_{cp}$ and LIV phase of $e \mu$ and $e \tau$ parameters can be observed as they appear in the appearance channel.
This discussion highlights the complex interplay between parameters in neutrino oscillation experiments and underscores the role of CP-violating phases and mass hierarchies in shaping experimental outcomes.

\section{Senstivity analysis}
\label{sec::sensitivity}
In this section, we discuss the methodological approach to study the effect of LIV parameters on the sensitivities of long-baseline experiment to determine, neutrino mass ordering and CP-violation by taking DUNE as a case study.

In order to derive the sensitivity, we adopt the Poisson-likelihood chi-square statistics. The Poisson-likelihood chi-square function for DUNE experiment can be  written as~\cite{Baker:1983tu}:
\setlength\abovedisplayskip{5pt}
\begin{widetext}
\begin{align}
\label{eq:chisq}
\Delta \chi^{2} = {\text{Min}} \Bigg[&2\sum_{x}^{\text{mode}}\sum_{j}^{\text{channel}}\sum_{i}^{\text{bin}}\Bigg\{
N_{ijx}^{\text{test}}(p^{\text{test}}) - N_{ijx}^{\text{true}}(p^{\text{true}})
+ N_{ijx}^{\text{true}}(p^{\text{true}}) \ln\frac{N_{ijx}^{\text{true}}(p^{\text{true}})}{N_{ijx}^{\text{test}}(p^{\text{test}})} \Bigg\}  \nonumber \\
\end{align}
\end{widetext}

In Equation \ref{eq:chisq}, $N^{\text{true}}$ and $N^{\text{test}}$ represent the sets of true and test events, respectively. The index $i$ sums over energy bins ranging from 0 to 20 GeV, totaling 71 bins within that range. Specifically, there are 64 bins with a width of $0.125$ GeV each from 0 to 8 GeV, and 7 bins with variable widths beyond 8 GeV. Events in each energy bin are summed over the total sidereal period, thus containing information of average effect of sidereal variation. For simplicity and lack of experimental information, exposure over the total sidereal period is considered to be constant. 
Indices $j$ and $x$ are summed over channels ($\nu_e, \nu_{\mu}$) and modes ($\nu$ and $\bar{\nu}$), respectively. 
The term inside the curly braces in Eq. \ref{eq:chisq} denotes the statistical component of $\chi^2$. The expression $(N^{\text{test}} - N^{\text{true}})$ captures the algebraic difference, 

The third term within the curly braces addresses the fractional difference between the true set and the test set of events. for true( $p^{\text{true}}$) and test( $p^{\text{test}}$) values of oscillation parameters . The True or best-fit values of oscillation parameters and corresponding uncertainties utilized in present analysis are detailed in table~\ref{table2OScPar}.

 Here the test parameteres may be some of  
$\ldm, \tb, \tc, \delcp, \phi^X_{\alpha \beta}, \phi^{XY}_{\alpha \beta}$ depending on the various chi square sensitivity scenarios.
We maintain a fixed total runtime of 10 years at DUNE, equally divided between neutrino and antineutrino modes.
\section{Results}
\label{sec::results}
In this section, We now examine how different LIV parameters effect
sensitivity of DUNE to determining the neutrino mass hierarchy (MH) and detecting CP violation. We further explore the precision of measurements related to the Dirac CP phase.
 \subsection{MH sensitivity} 
 Initially, we address Mass Hierarchy (MH) sensitivity, considering one of the mass hierarchy(NH/IH) as a true case and marginalizing the test parameter $\Delta m^2_{31}$ in the opposite hierarchy(IH/NH) within range as table~\ref{table2OScPar}. Concurrently, test parameters $\theta_{23}$ and $\theta_{13}$, are marginalized over their ranges shown in table~\ref{table2OScPar}, while the $\delcp$ is comprehensively marginalized over the entire $[-\pi, \pi]$ range. Additionally, LIV phases $\phi^X_{\alpha \beta}$ and $\phi^{XY}_{\alpha \beta}$ undergo marginalization over $[-\pi, \pi]$ in presence of corresponding partiular LIV parameter. 
 
In Figure~\ref{MHsensitivity}, we present a bi-sensitivity plot illustrating the MH-sensitivity of the DUNE experiment. The allowed sensitivity regions for the Standard Model (SM) and the combination of SM with Lorentz Invariance Violation (LIV) cases are depicted in black and red colors, respectively.
The X-axis in the figure represents the sensitivity of the experiment assuming the Normal Hierarchy (NH) as the true hierarchy, while the Y-axis represents the sensitivity assuming the Inverted Hierarchy (IH) as true scenario. The curve shows how sensitivity varies at different values of $\delcp$ in the range of [$\pi$, $-\pi$]. MH-sensitivity corresponding to the true $\delcp$ = $0$, $\pi/2$, $\pi$, $-\pi /2$ is shown by bullet points. 
DUNE is capable to determine the neutirno mass ordering in the SM case for the 10 year run of equal neutrino and anti-neutirno mode ~\cite{DUNE:2020jqi}.
The impact of $\cxyab$ type parameters is much stronger than $\axab$ type parameters, as their contribution increases linearly with energy.
The \( e\mu \) and \( e\tau \) type parameters exhibit minor sensitivity suppression but alter the dependency on \( \delcp\) due to their strong phase mixing with the standard CP-phase. In contrast, \( \mu\tau \) parametes show more significant suppression, particularly noticeable with parameter \( c^{XY}_{\mu\tau} \), where sensitivity is notably compromised in both the inverted hierarchy (IH) and normal hierarchy (NH) scenarios. Bi-probability plots (~\ref{biprobb}) indicate that the presence of \( c^{XY}_{\mu\tau} \) causes IH and NH regions to converge, potentially reducing sensitivity to mass hierarchy (MH) determination in DUNE.
Nevertheless, despite the presence of any of the considered parameters, sensitivity remains above the $5\sigma$ threshold.
In summary, the incorporation of these parameters, while impacting sensitivity, does not compromise the ability to achieve robust results within the DUNE experiment for the specified run period and parameter strength of the LIV parameter.
 \subsection{CP violation discovery potential}
 \label{secCPV}
	The determination of the CP-violating phase $\delta_{\text{CP}}$ stands as a formidable challenge within contemporary neutrino physics. In this section, we delve into the impact of Lorentz invariance violation (LIV) parameters on the CP violation sensitivity of the DUNE experiment. The significance of discerning CP violation, i.e., $\delta_{\text{CP}} = 0, or \pm \pi$, is illustrated in the accompanying figure ~\ref{Chi2delcpSensitivity}. In order to address the sensitivity of CP Violation, we restrict the test $\delta_{cp}$ to marginalize only over CP-conserving values of $0$ and $\pi$, allowing the true $\delta_{cp}$ to span the range $[-\pi, \pi]$. Simultaneously, test parameters $\Delta m^2_{31}$, $\theta_{23}$, and $\theta_{13}$ undergo marginalization over their respective marginalization ranges as shown in table ~\ref{table2OScPar}.
	
Figure~\ref{Chi2delcpSensitivity} depict the CP-violation (CPV) sensitivity of DUNE across various true values of $\delta_{\text{CP}}$, considering both neutrino and antineutrino channels. In each plot, the standard scenario is denoted by the black curve, while the red line denotes the presence of a LIV parameters. The corresponding test LIV phase parameter ($\phxab$/$\phxyab$) is marginalized in the range $[-\pi, \pi]$.
 In the SM scenario, DUNE demonstrates robust sensitivity in investigating CP violation over 10 year period with equal period each for neutrino and antineutrino modes ~\cite{DUNE:2020jqi}.
 The presence of the $a^X_{\alpha\beta}$ parameter, particularly $\axmt$, doesn't alter sensitivity significantly. 
  However, the inclusion of $\axem$, $\axet$, and $\cxymt$ parameters results in a slight decrease in sensitivity, albeit remaining at nearly $5\sigma$. Notably, a significant deterioration  in sensitivity arises from the presence of $\cxyem$ and $\cxyet$ parameters, reducing sensitivity to $2\sigma$. 
These points can be understood through two key factors. Firstly, the presence of \( e\mu \) and \( e\tau \) type parameters in the appearance channel, which is crucial for CP sensitivity. This results as a strong correlation between the $\phi^{X}_{e \mu}$ and $\phi^{XY}_{e \tau}$ with $\delcp$. Secondly, the strength of \( c^{XY}_{\alpha\beta} \) parameters increases with energy. With the low-energy beam flux of DUNE, where neutrino energies mainly range from 1 to 5 GeV, the contribution of \( c^{XY}_{\alpha\beta} \) from higher neutrino energies surpasses that of \( a^{XY}_{\alpha\beta} \). Therefore, the deterioration caused by \( c^{XY}_{\alpha\beta} \) parameters is more significant compared to \( a^{XY}_{\alpha\beta} \) parameters. It becomes evident that these energy-dependent \( c^{XY}_{\alpha\beta} \) parameters play a crucial role in compromising CP sensitivity overall.
	\begin{widetext}
	
	\begin{figure}[!h]
      \centering
       \includegraphics[height= 0.40\textwidth,width=0.90\textwidth]{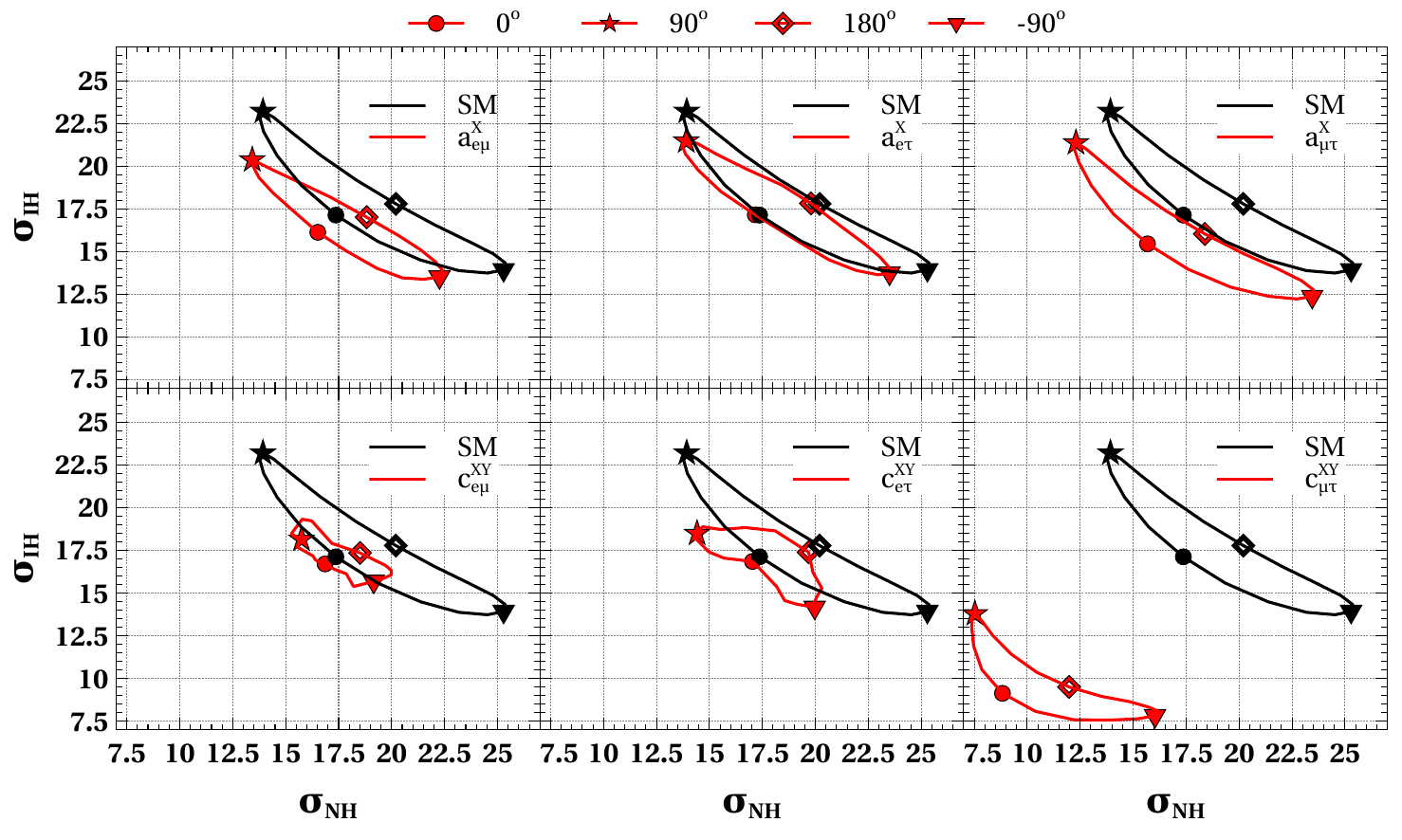}
      \caption{Bi-sensitivity plot illustrating the MH sensitivity in the case of DUNE. X-axis (Y-axis) corresponds to the sensitivity in the true NH(IH) for DUNE. Upper panel (Lower) corresponds to the $\axab$  ($\cxyab$). Left to right corresponds to the $e \mu, e \tau, \mu \tau$ type parameters.}
     \label{MHsensitivity}
    \end{figure} 
    	 \begin{figure}[!h]
      \centering
       \includegraphics[height= 0.40\textwidth,width=0.90\textwidth]{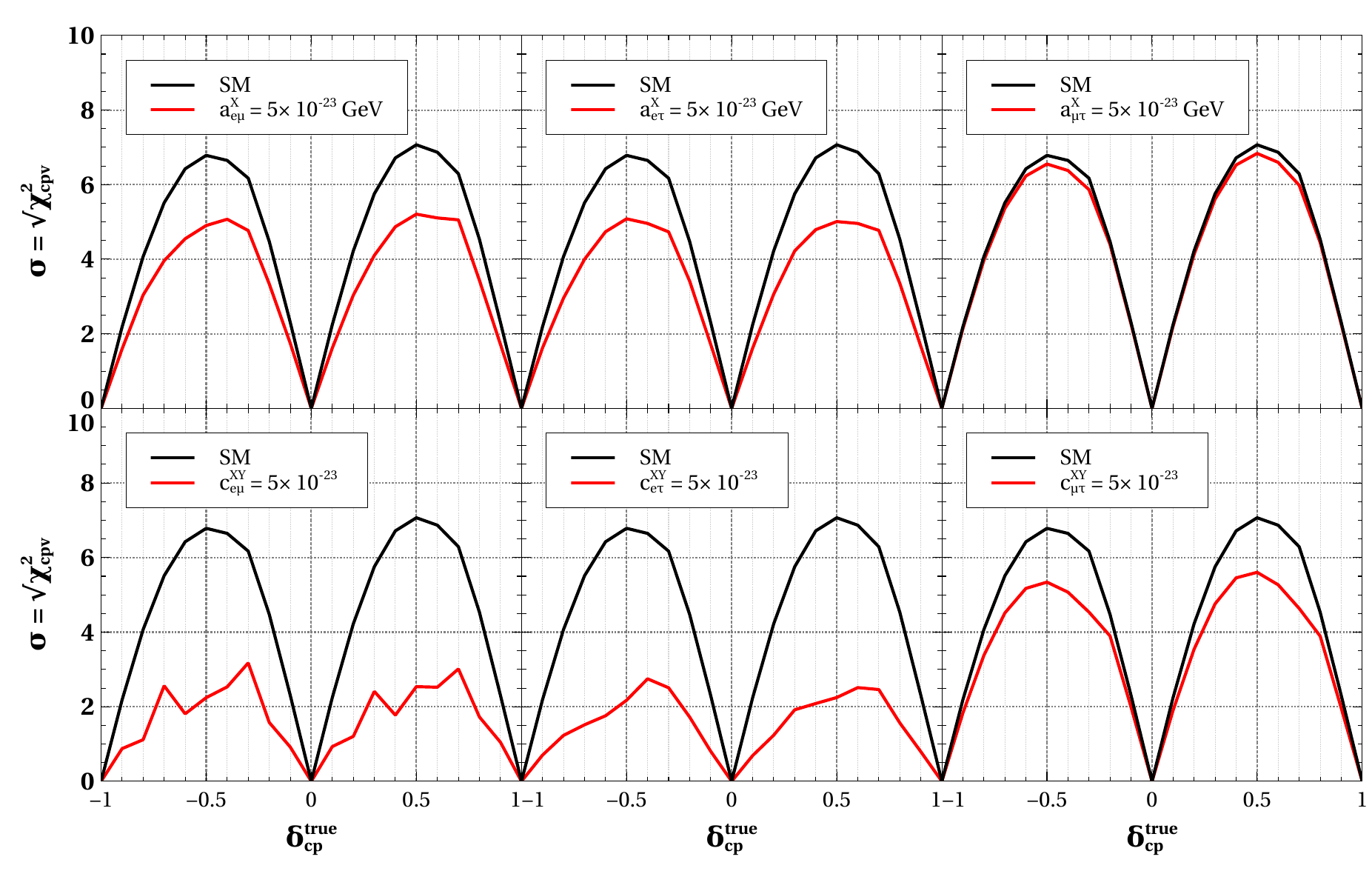}
      \caption{CP violation sensitivity as a function of the true values of \( \delta_{CP} \) for the DUNE experiment. The standard case is represented by the black curve in each plot. The red line indicates the presence of Lorentz invariance violation (LIV) parameters. The top (bottom) panel corresponds to \( a^{X}_{\alpha\beta} \) (\( c^{XY}_{\alpha\beta} \)) type parameters. From left to right, these parameters represent \( e\mu, e\tau, \mu\tau \) types.
 }
     \label{Chi2delcpSensitivity}
    \end{figure}
    
	\end{widetext}

	 \subsection{\bf{CP precision sesitivity}} 
	
\begin{widetext}
	
    \begin{figure}[H]
      \centering
       \includegraphics[height= 0.50\textwidth,width=0.75\textwidth]{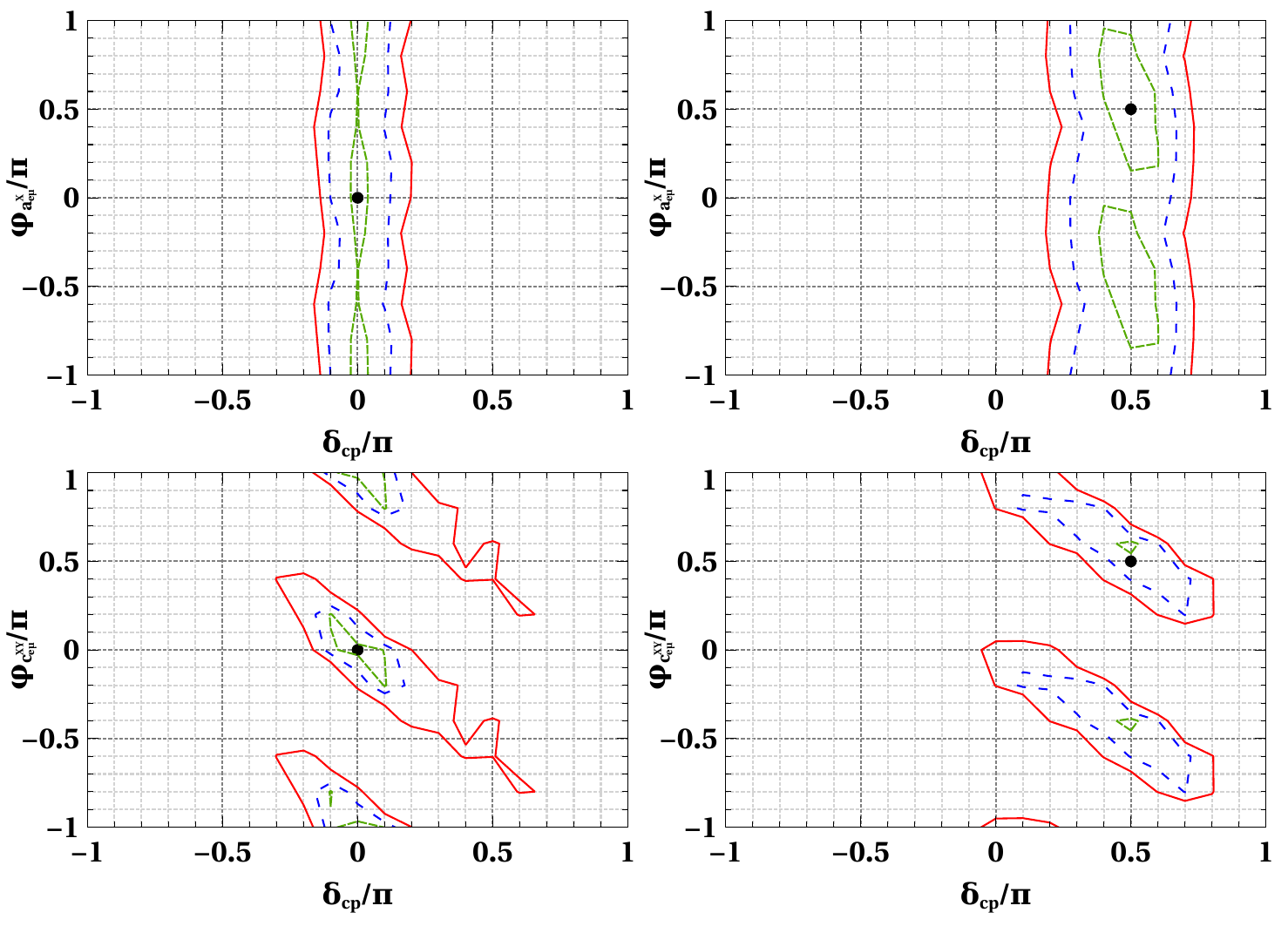}
      \caption{CP Precision sensitivity for two different true values of $\delcp$  and $\phi^{X}_{e \mu} (\phi^{XY}_{e \mu})$ for LIV parameter $\axem (\cxyem)$ for the DUNE configuration at Top (Bottom) panels. Red line, green line and dotted blue line represent the $1\sigma, 2\sigma$ and $3\sigma$ C.L. respectively.}
     \label{ReconstructDcp}
    \end{figure}
    
	\end{widetext}

	We further investigate the relationship between the phase of the LIV parameter and the testing of $\delta_{CP}$, a crucial aspect for accurately reconstructing the true $\delta_{CP}$ phases. To simplify matters, we focus on the $\axem$ and $\cxyem$ parameters\footnotemark.
	 \footnotetext{All $a^{X}_{\alpha \beta}$ type parameters have nearly the same reconstruction, similarly all $c^{XY}_{\alpha \beta}$ type parameters exhibit similar reconstruction}
	  Here, both the tested $\phi^X_{\alpha \beta}$ (or $\phi^{XY}_{\alpha \beta}$) and the tested $\delta_{cp}$ are allowed within specific combination, such as $[\pi/2, \pi/2]$, $[0, 0]$,  while parameters like $\Delta m^2_{31}$, $\theta_{23}$, $\theta_{13}$, and $\delta_{CP}$ undergo marginalization across their respective ranges as per table~\ref{table2OScPar}.
In Figure~\ref{ReconstructDcp}, we illustrate DUNE's capability to precisely reconstruct the CP-phase. The area beyond the $3\sigma$ contours indicates pairs of tested CP-phases that can be confidently excluded above $3\sigma$ when reconstructing their values for a specific choice of true CP phases. Smaller enclosed regions within the contours signify improved measurement accuracy. Notably, the reconstruction contours narrow down more effectively with the $\axem$ parameter compared to the $\cxyem$ parameter across all combinations.
	 \subsection{CP sensitivity with Exposure} 
 The analysis presented in Fig.~\ref{Chi2delcpSensitivity} underscores a notable decline in CP violation sensitivity due to the influence of a specific $\cxyab$ parameter. Given that one of the primary objectives of the DUNE project is the precise measurement of the CP-violating phase $\delta_{\text{CP}}$, it becomes imperative to assess the timeframe required to differentiate between scenarios where $\delta_{\text{CP}}$ equals 0 or $\pi/2$ within the Standard Model (SM), alongside scenarios incorporating all conceivable LIV cases.
 Furthermore, in assessing the total years required for achieving CPV sensitivity in the presence of a LIV parameter, true $\delta_{CP}$ are constrained to specific values ($0$) as CP conserving and ($\frac{\pi}{2}$) as CP violating. Test parameters $\Delta m^2_{31}$, $\theta_{23}$, $\theta_{13}$, and $\delta_{CP}$ are then marginalized over their respective marginalization ranges, alongside LIV phases $\phi^X_{\alpha \beta}$($\phi^{XY}_{\alpha \beta}$) over $[-\pi, \pi]$.
In the standard scenario, DUNE demonstrates the capacity to resolve the genuine CP-phase within a 5-year timeframe, split evenly between neutrino and antineutrino exposure, depicted by the black line. However, when considering LIV as a credible physics framework, the combination of LIV and CP-phases reduces DUNE's efficiency. Consequently, the duration necessary to probe the genuine CP phase also escalates. This increment has a milder impact on $\mu \tau$ type parameters, given their exclusive involvement in the disappearance channel, which exhibits lower sensitivity to CP. Conversely, the disparity becomes significantly pronounced for $e \mu$ and $e \tau$ type LIV parameters, as they directly influence the appearance channel, which is inherently sensitive to CP.
At the specified value of the $\cxymt$ parameter, DUNE's quest to discern $\delcp$ at a $3\sigma$ confidence level will extend beyond a decade. Furthermore, amplification of the $\cxymt$ parameter value may worsen this temporal requirement. 
	  \begin{figure}[H]
      \centering
       \includegraphics[height= 0.40\textwidth,width=0.40\textwidth]{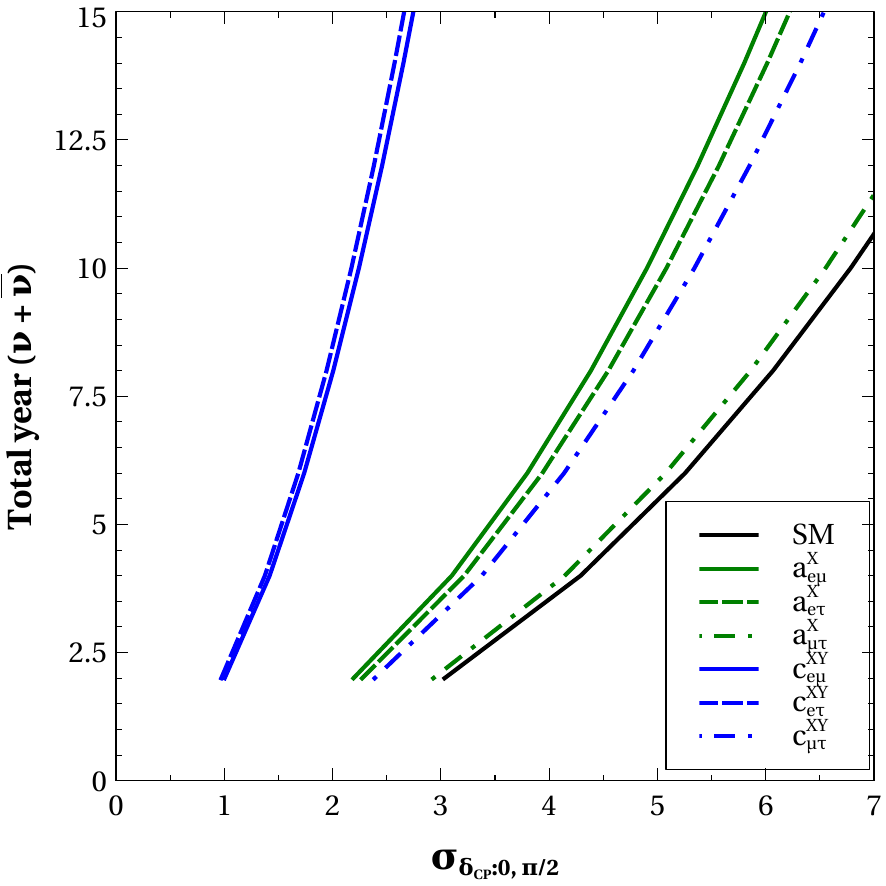}
      \caption{Total year required to distingush between $\delcp=0$ and $\delcp =\pi/2$ scenario in the SM and SM+LIV case for the DUNE is plotted. Total time frame has equal contribution from neutrino and antineutrino modes.}
    \end{figure}
         \section{Summary and conclusion}
\label{sec::summary}
 Lorentz Invariance is a cornerstone of spacetime symmetry, playing a pivotal role in our comprehension of neutrino oscillations' intricate details. Exploring potential deviations from this principle offers valuable insights into the nuanced variations that could affect neutrino oscillation probabilities. In contemporary neutrino physics, there is a strong emphasis on precision, with rigorous scrutiny applied to mixing parameters through diverse experimental approaches.

This study delves into the impact of Lorentz Invariance Violation (LIV) on the measurement capabilities of the forthcoming DUNE detector. Specifically, we focus on six non-isotropic LIV parameters denoted as $\axem, \axet, \axmt,\cxyem, \cxyet, \cxymt$, examining how they influence DUNE's ability to discern the standard unknowns of oscillation, particularly through the sidereal effect. 

For the MH sensitivity, $\cxyab$ type parameters have a stronger impact than  $\axab$ type parameters due to their linear energy dependence. $\axem, \axet, \cxyem$, and $\cxyet$ parameters show minor sensitivity suppression but affect the CP-phase dependence, while $c^{XY}_{\mu\tau}$, experience significant suppression, compromising sensitivity in both hierarchy scenarios.
 Bi-probability plots illustrate that \( C_{\mu\tau} \) causes convergence between IH and NH regions, potentially impacting MH determination in DUNE. Nevertheless, despite these effects, sensitivity remains consistently above the \( 5\sigma \) threshold.

The inclusion of parameters such as $\axem$, $\axet$, $\axmt$  and $\cxymt$ slightly decreases CP sensitivity although it remains close to $5\sigma$. However, the presence of $\cxyem$ and $\cxyet$ parameters notably reduces sensitivity to $2\sigma$, indicating their significant impact. Energy-dependent $c^{XY}_{\alpha\beta}$ parameters play a crucial role in compromising CP overall sensitivity. DUNE's capability to precisely reconstruct the CP-phase is also examined. Notably, the reconstruction contours narrow down more effectively with the $\axet$ parameter compared to the $\cxyet$ parameter across all combinations.

The study emphasizes the need for careful consideration of non-isotropic LIV parameter impacts and highlights areas where experimental sensitivity and precision can be affected. 

\section{Acknowledgments}
S. M., S. S. and V. S. acknowledge the Department of Science and Technology (DST), New Delhi, India, under the Umbrella Scheme for Research and Development. V. S. and L. S. acknowledge FIST-Department of Science and Technology (DST), New Delhi, India as being part of the Department of Physics, CUSB, Gaya, India. S. S. acknowledges financial support from the Council of Scientific and Industrial Research (CSIR), New Delhi, India. L. S. acknowledges support from the University Grants Commission--Basic Scientific Research Faculty Fellowship Scheme (UGC-BSR) Research Start Up Grant, India (Contract No. F.30-584/2021 (BSR)).
     
\bibliography{DUNEcpv}

\begin{thebibliography}{40}%
\makeatletter
\providecommand \@ifxundefined [1]{%
 \@ifx{#1\undefined}
}%
\providecommand \@ifnum [1]{%
 \ifnum #1\expandafter \@firstoftwo
 \else \expandafter \@secondoftwo
 \fi
}%
\providecommand \@ifx [1]{%
 \ifx #1\expandafter \@firstoftwo
 \else \expandafter \@secondoftwo
 \fi
}%
\providecommand \natexlab [1]{#1}%
\providecommand \enquote  [1]{``#1''}%
\providecommand \bibnamefont  [1]{#1}%
\providecommand \bibfnamefont [1]{#1}%
\providecommand \citenamefont [1]{#1}%
\providecommand \href@noop [0]{\@secondoftwo}%
\providecommand \href [0]{\begingroup \@sanitize@url \@href}%
\providecommand \@href[1]{\@@startlink{#1}\@@href}%
\providecommand \@@href[1]{\endgroup#1\@@endlink}%
\providecommand \@sanitize@url [0]{\catcode `\\12\catcode `\$12\catcode
  `\&12\catcode `\#12\catcode `\^12\catcode `\_12\catcode `\%12\relax}%
\providecommand \@@startlink[1]{}%
\providecommand \@@endlink[0]{}%
\providecommand \url  [0]{\begingroup\@sanitize@url \@url }%
\providecommand \@url [1]{\endgroup\@href {#1}{\urlprefix }}%
\providecommand \urlprefix  [0]{URL }%
\providecommand \Eprint [0]{\href }%
\providecommand \doibase [0]{https://doi.org/}%
\providecommand \selectlanguage [0]{\@gobble}%
\providecommand \bibinfo  [0]{\@secondoftwo}%
\providecommand \bibfield  [0]{\@secondoftwo}%
\providecommand \translation [1]{[#1]}%
\providecommand \BibitemOpen [0]{}%
\providecommand \bibitemStop [0]{}%
\providecommand \bibitemNoStop [0]{.\EOS\space}%
\providecommand \EOS [0]{\spacefactor3000\relax}%
\providecommand \BibitemShut  [1]{\csname bibitem#1\endcsname}%
\let\auto@bib@innerbib\@empty
\bibitem [{\citenamefont {Workman}\ and\ \citenamefont
  {Others}(2022)}]{Workman:2022ynf}%
  \BibitemOpen
  \bibfield  {author} {\bibinfo {author} {\bibfnamefont {R.~L.}\ \bibnamefont
  {Workman}}\ and\ \bibinfo {author} {\bibnamefont {Others}} (\bibinfo
  {collaboration} {Particle Data Group}),\ }\bibfield  {title} {\bibinfo
  {title} {{Review of Particle Physics}},\ }\href
  {https://doi.org/10.1093/ptep/ptac097} {\bibfield  {journal} {\bibinfo
  {journal} {PTEP}\ }\textbf {\bibinfo {volume} {2022}},\ \bibinfo {pages}
  {083C01} (\bibinfo {year} {2022})}\BibitemShut {NoStop}%
\bibitem [{\citenamefont {Esteban}\ \emph {et~al.}(2020)\citenamefont
  {Esteban}, \citenamefont {Gonzalez-Garcia}, \citenamefont {Maltoni},
  \citenamefont {Schwetz},\ and\ \citenamefont {Zhou}}]{Esteban:2020cvm}%
  \BibitemOpen
  \bibfield  {author} {\bibinfo {author} {\bibfnamefont {I.}~\bibnamefont
  {Esteban}}, \bibinfo {author} {\bibfnamefont {M.~C.}\ \bibnamefont
  {Gonzalez-Garcia}}, \bibinfo {author} {\bibfnamefont {M.}~\bibnamefont
  {Maltoni}}, \bibinfo {author} {\bibfnamefont {T.}~\bibnamefont {Schwetz}},\
  and\ \bibinfo {author} {\bibfnamefont {A.}~\bibnamefont {Zhou}},\ }\bibfield
  {title} {\bibinfo {title} {{The fate of hints: updated global analysis of
  three-flavor neutrino oscillations}},\ }\href
  {https://doi.org/10.1007/JHEP09(2020)178} {\bibfield  {journal} {\bibinfo
  {journal} {JHEP}\ }\textbf {\bibinfo {volume} {09}},\ \bibinfo {pages}
  {178}},\ \Eprint {https://arxiv.org/abs/2007.14792} {arXiv:2007.14792
  [hep-ph]} \BibitemShut {NoStop}%
\bibitem [{\citenamefont {Fukugita}\ and\ \citenamefont
  {Yanagida}(1986)}]{FUKUGITA198645}%
  \BibitemOpen
  \bibfield  {author} {\bibinfo {author} {\bibfnamefont {M.}~\bibnamefont
  {Fukugita}}\ and\ \bibinfo {author} {\bibfnamefont {T.}~\bibnamefont
  {Yanagida}},\ }\bibfield  {title} {\bibinfo {title} {{Baryogenesis Without
  Grand Unification}},\ }\href {https://doi.org/10.1016/0370-2693(86)91126-3}
  {\bibfield  {journal} {\bibinfo  {journal} {Phys. Lett. B}\ }\textbf
  {\bibinfo {volume} {174}},\ \bibinfo {pages} {45} (\bibinfo {year}
  {1986})}\BibitemShut {NoStop}%
\bibitem [{\citenamefont {Branco}\ \emph {et~al.}(2012)\citenamefont {Branco},
  \citenamefont {Felipe},\ and\ \citenamefont {Joaquim}}]{RevModPhys.84.515}%
  \BibitemOpen
  \bibfield  {author} {\bibinfo {author} {\bibfnamefont {G.~C.}\ \bibnamefont
  {Branco}}, \bibinfo {author} {\bibfnamefont {R.~G.}\ \bibnamefont {Felipe}},\
  and\ \bibinfo {author} {\bibfnamefont {F.~R.}\ \bibnamefont {Joaquim}},\
  }\bibfield  {title} {\bibinfo {title} {{Leptonic CP Violation}},\ }\href
  {https://doi.org/10.1103/RevModPhys.84.515} {\bibfield  {journal} {\bibinfo
  {journal} {Rev. Mod. Phys.}\ }\textbf {\bibinfo {volume} {84}},\ \bibinfo
  {pages} {515} (\bibinfo {year} {2012})},\ \Eprint
  {https://arxiv.org/abs/1111.5332} {arXiv:1111.5332 [hep-ph]} \BibitemShut
  {NoStop}%
\bibitem [{\citenamefont {Haxton}\ and\ \citenamefont
  {Stephenson}(1984)}]{HAXTON1984409}%
  \BibitemOpen
  \bibfield  {author} {\bibinfo {author} {\bibfnamefont {W.~C.}\ \bibnamefont
  {Haxton}}\ and\ \bibinfo {author} {\bibfnamefont {G.~J.}\ \bibnamefont
  {Stephenson}},\ }\bibfield  {title} {\bibinfo {title} {{Double beta Decay}},\
  }\href {https://doi.org/10.1016/0146-6410(84)90006-1} {\bibfield  {journal}
  {\bibinfo  {journal} {Prog. Part. Nucl. Phys.}\ }\textbf {\bibinfo {volume}
  {12}},\ \bibinfo {pages} {409} (\bibinfo {year} {1984})}\BibitemShut
  {NoStop}%
\bibitem [{\citenamefont {Fiza}\ \emph {et~al.}(2023)\citenamefont {Fiza},
  \citenamefont {Khan~Chowdhury},\ and\ \citenamefont {Masud}}]{Fiza:2022xfw}%
  \BibitemOpen
  \bibfield  {author} {\bibinfo {author} {\bibfnamefont {N.}~\bibnamefont
  {Fiza}}, \bibinfo {author} {\bibfnamefont {N.~R.}\ \bibnamefont
  {Khan~Chowdhury}},\ and\ \bibinfo {author} {\bibfnamefont {M.}~\bibnamefont
  {Masud}},\ }\bibfield  {title} {\bibinfo {title} {{Investigating Lorentz
  Invariance Violation with the long baseline experiment P2O}},\ }\href
  {https://doi.org/10.1007/JHEP01(2023)076} {\bibfield  {journal} {\bibinfo
  {journal} {JHEP}\ }\textbf {\bibinfo {volume} {01}},\ \bibinfo {pages}
  {076}},\ \Eprint {https://arxiv.org/abs/2206.14018} {arXiv:2206.14018
  [hep-ph]} \BibitemShut {NoStop}%
\bibitem [{\citenamefont {Kumar~Agarwalla}\ and\ \citenamefont
  {Masud}(2020)}]{KumarAgarwalla:2019gdj}%
  \BibitemOpen
  \bibfield  {author} {\bibinfo {author} {\bibfnamefont {S.}~\bibnamefont
  {Kumar~Agarwalla}}\ and\ \bibinfo {author} {\bibfnamefont {M.}~\bibnamefont
  {Masud}},\ }\bibfield  {title} {\bibinfo {title} {{Can Lorentz invariance
  violation affect the sensitivity of deep underground neutrino experiment?}},\
  }\href {https://doi.org/10.1140/epjc/s10052-020-8303-1} {\bibfield  {journal}
  {\bibinfo  {journal} {Eur. Phys. J. C}\ }\textbf {\bibinfo {volume} {80}},\
  \bibinfo {pages} {716} (\bibinfo {year} {2020})},\ \Eprint
  {https://arxiv.org/abs/1912.13306} {arXiv:1912.13306 [hep-ph]} \BibitemShut
  {NoStop}%
\bibitem [{\citenamefont {Masud}\ and\ \citenamefont
  {Mehta}(2016{\natexlab{a}})}]{Masud:2016nuj}%
  \BibitemOpen
  \bibfield  {author} {\bibinfo {author} {\bibfnamefont {M.}~\bibnamefont
  {Masud}}\ and\ \bibinfo {author} {\bibfnamefont {P.}~\bibnamefont {Mehta}},\
  }\bibfield  {title} {\bibinfo {title} {{Nonstandard interactions and
  resolving the ordering of neutrino masses at DUNE and other long baseline
  experiments}},\ }\href {https://doi.org/10.1103/PhysRevD.94.053007}
  {\bibfield  {journal} {\bibinfo  {journal} {Phys. Rev. D}\ }\textbf {\bibinfo
  {volume} {94}},\ \bibinfo {pages} {053007} (\bibinfo {year}
  {2016}{\natexlab{a}})},\ \Eprint {https://arxiv.org/abs/1606.05662}
  {arXiv:1606.05662 [hep-ph]} \BibitemShut {NoStop}%
\bibitem [{\citenamefont {Masud}\ and\ \citenamefont
  {Mehta}(2016{\natexlab{b}})}]{Masud:2016bvp}%
  \BibitemOpen
  \bibfield  {author} {\bibinfo {author} {\bibfnamefont {M.}~\bibnamefont
  {Masud}}\ and\ \bibinfo {author} {\bibfnamefont {P.}~\bibnamefont {Mehta}},\
  }\bibfield  {title} {\bibinfo {title} {{Nonstandard interactions spoiling the
  CP violation sensitivity at DUNE and other long baseline experiments}},\
  }\href {https://doi.org/10.1103/PhysRevD.94.013014} {\bibfield  {journal}
  {\bibinfo  {journal} {Phys. Rev. D}\ }\textbf {\bibinfo {volume} {94}},\
  \bibinfo {pages} {013014} (\bibinfo {year} {2016}{\natexlab{b}})},\ \Eprint
  {https://arxiv.org/abs/1603.01380} {arXiv:1603.01380 [hep-ph]} \BibitemShut
  {NoStop}%
\bibitem [{\citenamefont {Pan}\ \emph {et~al.}(2024)\citenamefont {Pan},
  \citenamefont {Chakraborty},\ and\ \citenamefont {Goswami}}]{Pan:2023qln}%
  \BibitemOpen
  \bibfield  {author} {\bibinfo {author} {\bibfnamefont {S.}~\bibnamefont
  {Pan}}, \bibinfo {author} {\bibfnamefont {K.}~\bibnamefont {Chakraborty}},\
  and\ \bibinfo {author} {\bibfnamefont {S.}~\bibnamefont {Goswami}},\
  }\bibfield  {title} {\bibinfo {title} {{Sensitivity to CP discovery in the
  presence of Lorentz invariance-violating potential at T2HK/T2HKK}},\ }\href
  {https://doi.org/10.1140/epjc/s10052-024-12541-y} {\bibfield  {journal}
  {\bibinfo  {journal} {Eur. Phys. J. C}\ }\textbf {\bibinfo {volume} {84}},\
  \bibinfo {pages} {354} (\bibinfo {year} {2024})},\ \Eprint
  {https://arxiv.org/abs/2308.07566} {arXiv:2308.07566 [hep-ph]} \BibitemShut
  {NoStop}%
\bibitem [{\citenamefont {Raikwal}\ \emph {et~al.}(2023)\citenamefont
  {Raikwal}, \citenamefont {Choubey},\ and\ \citenamefont
  {Ghosh}}]{Raikwal:2023lzk}%
  \BibitemOpen
  \bibfield  {author} {\bibinfo {author} {\bibfnamefont {D.}~\bibnamefont
  {Raikwal}}, \bibinfo {author} {\bibfnamefont {S.}~\bibnamefont {Choubey}},\
  and\ \bibinfo {author} {\bibfnamefont {M.}~\bibnamefont {Ghosh}},\ }\bibfield
   {title} {\bibinfo {title} {{Comprehensive study of Lorentz invariance
  violation in atmospheric and long-baseline experiments}},\ }\href
  {https://doi.org/10.1103/PhysRevD.107.115032} {\bibfield  {journal} {\bibinfo
   {journal} {Phys. Rev. D}\ }\textbf {\bibinfo {volume} {107}},\ \bibinfo
  {pages} {115032} (\bibinfo {year} {2023})},\ \Eprint
  {https://arxiv.org/abs/2303.10892} {arXiv:2303.10892 [hep-ph]} \BibitemShut
  {NoStop}%
\bibitem [{\citenamefont {Kostelecky}\ and\ \citenamefont
  {Samuel}(1989)}]{Kostelecky:1988zi}%
  \BibitemOpen
  \bibfield  {author} {\bibinfo {author} {\bibfnamefont {V.~A.}\ \bibnamefont
  {Kostelecky}}\ and\ \bibinfo {author} {\bibfnamefont {S.}~\bibnamefont
  {Samuel}},\ }\bibfield  {title} {\bibinfo {title} {{Spontaneous Breaking of
  Lorentz Symmetry in String Theory}},\ }\href
  {https://doi.org/10.1103/PhysRevD.39.683} {\bibfield  {journal} {\bibinfo
  {journal} {Phys. Rev. D}\ }\textbf {\bibinfo {volume} {39}},\ \bibinfo
  {pages} {683} (\bibinfo {year} {1989})}\BibitemShut {NoStop}%
\bibitem [{\citenamefont {Kostelecky}\ and\ \citenamefont
  {Potting}(1991)}]{Kostelecky:1991ak}%
  \BibitemOpen
  \bibfield  {author} {\bibinfo {author} {\bibfnamefont {V.~A.}\ \bibnamefont
  {Kostelecky}}\ and\ \bibinfo {author} {\bibfnamefont {R.}~\bibnamefont
  {Potting}},\ }\bibfield  {title} {\bibinfo {title} {{CPT and strings}},\
  }\href {https://doi.org/10.1016/0550-3213(91)90071-5} {\bibfield  {journal}
  {\bibinfo  {journal} {Nucl. Phys. B}\ }\textbf {\bibinfo {volume} {359}},\
  \bibinfo {pages} {545} (\bibinfo {year} {1991})}\BibitemShut {NoStop}%
\bibitem [{\citenamefont {Colladay}\ and\ \citenamefont
  {Kostelecky}(1998)}]{Colladay:1998fq}%
  \BibitemOpen
  \bibfield  {author} {\bibinfo {author} {\bibfnamefont {D.}~\bibnamefont
  {Colladay}}\ and\ \bibinfo {author} {\bibfnamefont {V.~A.}\ \bibnamefont
  {Kostelecky}},\ }\bibfield  {title} {\bibinfo {title} {{Lorentz violating
  extension of the standard model}},\ }\href
  {https://doi.org/10.1103/PhysRevD.58.116002} {\bibfield  {journal} {\bibinfo
  {journal} {Phys. Rev. D}\ }\textbf {\bibinfo {volume} {58}},\ \bibinfo
  {pages} {116002} (\bibinfo {year} {1998})},\ \Eprint
  {https://arxiv.org/abs/hep-ph/9809521} {arXiv:hep-ph/9809521} \BibitemShut
  {NoStop}%
\bibitem [{\citenamefont {Bluhm}(2006)}]{Bluhm:2005uj}%
  \BibitemOpen
  \bibfield  {author} {\bibinfo {author} {\bibfnamefont {R.}~\bibnamefont
  {Bluhm}},\ }\bibfield  {title} {\bibinfo {title} {{Overview of the SME:
  Implications and phenomenology of Lorentz violation}},\ }\href
  {https://doi.org/10.1007/3-540-34523-X_8} {\bibfield  {journal} {\bibinfo
  {journal} {Lect. Notes Phys.}\ }\textbf {\bibinfo {volume} {702}},\ \bibinfo
  {pages} {191} (\bibinfo {year} {2006})},\ \Eprint
  {https://arxiv.org/abs/hep-ph/0506054} {arXiv:hep-ph/0506054} \BibitemShut
  {NoStop}%
\bibitem [{\citenamefont {Diaz}\ \emph {et~al.}(2009)\citenamefont {Diaz},
  \citenamefont {Kostelecky},\ and\ \citenamefont {Mewes}}]{Diaz:2009qk}%
  \BibitemOpen
  \bibfield  {author} {\bibinfo {author} {\bibfnamefont {J.~S.}\ \bibnamefont
  {Diaz}}, \bibinfo {author} {\bibfnamefont {V.~A.}\ \bibnamefont
  {Kostelecky}},\ and\ \bibinfo {author} {\bibfnamefont {M.}~\bibnamefont
  {Mewes}},\ }\bibfield  {title} {\bibinfo {title} {{Perturbative Lorentz and
  CPT violation for neutrino and antineutrino oscillations}},\ }\href
  {https://doi.org/10.1103/PhysRevD.80.076007} {\bibfield  {journal} {\bibinfo
  {journal} {Phys. Rev. D}\ }\textbf {\bibinfo {volume} {80}},\ \bibinfo
  {pages} {076007} (\bibinfo {year} {2009})},\ \Eprint
  {https://arxiv.org/abs/0908.1401} {arXiv:0908.1401 [hep-ph]} \BibitemShut
  {NoStop}%
\bibitem [{\citenamefont {Kostelecky}\ and\ \citenamefont
  {Mewes}(2004{\natexlab{a}})}]{Kostelecky:2003xn}%
  \BibitemOpen
  \bibfield  {author} {\bibinfo {author} {\bibfnamefont {V.~A.}\ \bibnamefont
  {Kostelecky}}\ and\ \bibinfo {author} {\bibfnamefont {M.}~\bibnamefont
  {Mewes}},\ }\bibfield  {title} {\bibinfo {title} {{Lorentz and CPT violation
  in the neutrino sector}},\ }\href
  {https://doi.org/10.1103/PhysRevD.70.031902} {\bibfield  {journal} {\bibinfo
  {journal} {Phys. Rev. D}\ }\textbf {\bibinfo {volume} {70}},\ \bibinfo
  {pages} {031902} (\bibinfo {year} {2004}{\natexlab{a}})},\ \Eprint
  {https://arxiv.org/abs/hep-ph/0308300} {arXiv:hep-ph/0308300} \BibitemShut
  {NoStop}%
\bibitem [{\citenamefont {Kostelecky}\ and\ \citenamefont
  {Mewes}(2004{\natexlab{b}})}]{Kostelecky:2004hg}%
  \BibitemOpen
  \bibfield  {author} {\bibinfo {author} {\bibfnamefont {V.~A.}\ \bibnamefont
  {Kostelecky}}\ and\ \bibinfo {author} {\bibfnamefont {M.}~\bibnamefont
  {Mewes}},\ }\bibfield  {title} {\bibinfo {title} {{Lorentz violation and
  short-baseline neutrino experiments}},\ }\href
  {https://doi.org/10.1103/PhysRevD.70.076002} {\bibfield  {journal} {\bibinfo
  {journal} {Phys. Rev. D}\ }\textbf {\bibinfo {volume} {70}},\ \bibinfo
  {pages} {076002} (\bibinfo {year} {2004}{\natexlab{b}})},\ \Eprint
  {https://arxiv.org/abs/hep-ph/0406255} {arXiv:hep-ph/0406255} \BibitemShut
  {NoStop}%
\bibitem [{\citenamefont {Adamson}\ \emph {et~al.}(2008)\citenamefont {Adamson}
  \emph {et~al.}}]{MINOS:2008fnv}%
  \BibitemOpen
  \bibfield  {author} {\bibinfo {author} {\bibfnamefont {P.}~\bibnamefont
  {Adamson}} \emph {et~al.} (\bibinfo {collaboration} {MINOS}),\ }\bibfield
  {title} {\bibinfo {title} {{Testing Lorentz Invariance and CPT Conservation
  with NuMI Neutrinos in the MINOS Near Detector}},\ }\href
  {https://doi.org/10.1103/PhysRevLett.101.151601} {\bibfield  {journal}
  {\bibinfo  {journal} {Phys. Rev. Lett.}\ }\textbf {\bibinfo {volume} {101}},\
  \bibinfo {pages} {151601} (\bibinfo {year} {2008})},\ \Eprint
  {https://arxiv.org/abs/0806.4945} {arXiv:0806.4945 [hep-ex]} \BibitemShut
  {NoStop}%
\bibitem [{\citenamefont {Abe}\ \emph {et~al.}(2015)\citenamefont {Abe} \emph
  {et~al.}}]{Super-Kamiokande:2014exs}%
  \BibitemOpen
  \bibfield  {author} {\bibinfo {author} {\bibfnamefont {K.}~\bibnamefont
  {Abe}} \emph {et~al.} (\bibinfo {collaboration} {Super-Kamiokande}),\
  }\bibfield  {title} {\bibinfo {title} {{Test of Lorentz invariance with
  atmospheric neutrinos}},\ }\href {https://doi.org/10.1103/PhysRevD.91.052003}
  {\bibfield  {journal} {\bibinfo  {journal} {Phys. Rev. D}\ }\textbf {\bibinfo
  {volume} {91}},\ \bibinfo {pages} {052003} (\bibinfo {year} {2015})},\
  \Eprint {https://arxiv.org/abs/1410.4267} {arXiv:1410.4267 [hep-ex]}
  \BibitemShut {NoStop}%
\bibitem [{\citenamefont {Auerbach}\ \emph {et~al.}(2005)\citenamefont
  {Auerbach} \emph {et~al.}}]{LSND:2005oop}%
  \BibitemOpen
  \bibfield  {author} {\bibinfo {author} {\bibfnamefont {L.~B.}\ \bibnamefont
  {Auerbach}} \emph {et~al.} (\bibinfo {collaboration} {LSND}),\ }\bibfield
  {title} {\bibinfo {title} {{Tests of Lorentz violation in anti-nu(mu)
  ---\ensuremath{>} anti-nu(e) oscillations}},\ }\href
  {https://doi.org/10.1103/PhysRevD.72.076004} {\bibfield  {journal} {\bibinfo
  {journal} {Phys. Rev. D}\ }\textbf {\bibinfo {volume} {72}},\ \bibinfo
  {pages} {076004} (\bibinfo {year} {2005})},\ \Eprint
  {https://arxiv.org/abs/hep-ex/0506067} {arXiv:hep-ex/0506067} \BibitemShut
  {NoStop}%
\bibitem [{\citenamefont {Adamson}\ \emph {et~al.}(2010)\citenamefont {Adamson}
  \emph {et~al.}}]{MINOS:2010kat}%
  \BibitemOpen
  \bibfield  {author} {\bibinfo {author} {\bibfnamefont {P.}~\bibnamefont
  {Adamson}} \emph {et~al.} (\bibinfo {collaboration} {MINOS}),\ }\bibfield
  {title} {\bibinfo {title} {{A Search for Lorentz Invariance and CPT Violation
  with the MINOS Far Detector}},\ }\href
  {https://doi.org/10.1103/PhysRevLett.105.151601} {\bibfield  {journal}
  {\bibinfo  {journal} {Phys. Rev. Lett.}\ }\textbf {\bibinfo {volume} {105}},\
  \bibinfo {pages} {151601} (\bibinfo {year} {2010})},\ \Eprint
  {https://arxiv.org/abs/1007.2791} {arXiv:1007.2791 [hep-ex]} \BibitemShut
  {NoStop}%
\bibitem [{\citenamefont {Adamson}\ \emph {et~al.}(2012)\citenamefont {Adamson}
  \emph {et~al.}}]{MINOS:2012ozn}%
  \BibitemOpen
  \bibfield  {author} {\bibinfo {author} {\bibfnamefont {P.}~\bibnamefont
  {Adamson}} \emph {et~al.} (\bibinfo {collaboration} {MINOS}),\ }\bibfield
  {title} {\bibinfo {title} {{Search for Lorentz invariance and CPT violation
  with muon antineutrinos in the MINOS Near Detector}},\ }\href
  {https://doi.org/10.1103/PhysRevD.85.031101} {\bibfield  {journal} {\bibinfo
  {journal} {Phys. Rev. D}\ }\textbf {\bibinfo {volume} {85}},\ \bibinfo
  {pages} {031101} (\bibinfo {year} {2012})},\ \Eprint
  {https://arxiv.org/abs/1201.2631} {arXiv:1201.2631 [hep-ex]} \BibitemShut
  {NoStop}%
\bibitem [{\citenamefont {Aguilar-Arevalo}\ \emph {et~al.}(2013)\citenamefont
  {Aguilar-Arevalo} \emph {et~al.}}]{MiniBooNE:2011pix}%
  \BibitemOpen
  \bibfield  {author} {\bibinfo {author} {\bibfnamefont {A.~A.}\ \bibnamefont
  {Aguilar-Arevalo}} \emph {et~al.} (\bibinfo {collaboration} {MiniBooNE}),\
  }\bibfield  {title} {\bibinfo {title} {{Test of Lorentz and CPT violation
  with Short Baseline Neutrino Oscillation Excesses}},\ }\href
  {https://doi.org/10.1016/j.physletb.2012.12.020} {\bibfield  {journal}
  {\bibinfo  {journal} {Phys. Lett. B}\ }\textbf {\bibinfo {volume} {718}},\
  \bibinfo {pages} {1303} (\bibinfo {year} {2013})},\ \Eprint
  {https://arxiv.org/abs/1109.3480} {arXiv:1109.3480 [hep-ex]} \BibitemShut
  {NoStop}%
\bibitem [{\citenamefont {Abbasi}\ \emph {et~al.}(2010)\citenamefont {Abbasi}
  \emph {et~al.}}]{IceCube:2010fyu}%
  \BibitemOpen
  \bibfield  {author} {\bibinfo {author} {\bibfnamefont {R.}~\bibnamefont
  {Abbasi}} \emph {et~al.} (\bibinfo {collaboration} {IceCube}),\ }\bibfield
  {title} {\bibinfo {title} {{Search for a Lorentz-violating sidereal signal
  with atmospheric neutrinos in IceCube}},\ }\href
  {https://doi.org/10.1103/PhysRevD.82.112003} {\bibfield  {journal} {\bibinfo
  {journal} {Phys. Rev. D}\ }\textbf {\bibinfo {volume} {82}},\ \bibinfo
  {pages} {112003} (\bibinfo {year} {2010})},\ \Eprint
  {https://arxiv.org/abs/1010.4096} {arXiv:1010.4096 [astro-ph.HE]}
  \BibitemShut {NoStop}%
\bibitem [{\citenamefont {Abe}\ \emph {et~al.}(2012)\citenamefont {Abe} \emph
  {et~al.}}]{DoubleChooz:2012eiq}%
  \BibitemOpen
  \bibfield  {author} {\bibinfo {author} {\bibfnamefont {Y.}~\bibnamefont
  {Abe}} \emph {et~al.} (\bibinfo {collaboration} {Double Chooz}),\ }\bibfield
  {title} {\bibinfo {title} {{First Test of Lorentz Violation with a
  Reactor-based Antineutrino Experiment}},\ }\href
  {https://doi.org/10.1103/PhysRevD.86.112009} {\bibfield  {journal} {\bibinfo
  {journal} {Phys. Rev. D}\ }\textbf {\bibinfo {volume} {86}},\ \bibinfo
  {pages} {112009} (\bibinfo {year} {2012})},\ \Eprint
  {https://arxiv.org/abs/1209.5810} {arXiv:1209.5810 [hep-ex]} \BibitemShut
  {NoStop}%
\bibitem [{\citenamefont {Abe}\ \emph {et~al.}(2017)\citenamefont {Abe} \emph
  {et~al.}}]{T2K:2017ega}%
  \BibitemOpen
  \bibfield  {author} {\bibinfo {author} {\bibfnamefont {K.}~\bibnamefont
  {Abe}} \emph {et~al.} (\bibinfo {collaboration} {T2K}),\ }\bibfield  {title}
  {\bibinfo {title} {{Search for Lorentz and CPT violation using sidereal time
  dependence of neutrino flavor transitions over a short baseline}},\ }\href
  {https://doi.org/10.1103/PhysRevD.95.111101} {\bibfield  {journal} {\bibinfo
  {journal} {Phys. Rev. D}\ }\textbf {\bibinfo {volume} {95}},\ \bibinfo
  {pages} {111101} (\bibinfo {year} {2017})},\ \Eprint
  {https://arxiv.org/abs/1703.01361} {arXiv:1703.01361 [hep-ex]} \BibitemShut
  {NoStop}%
\bibitem [{\citenamefont {Abi}\ \emph {et~al.}(2020{\natexlab{a}})\citenamefont
  {Abi} \emph {et~al.}}]{DUNE:2020lwj}%
  \BibitemOpen
  \bibfield  {author} {\bibinfo {author} {\bibfnamefont {B.}~\bibnamefont
  {Abi}} \emph {et~al.} (\bibinfo {collaboration} {DUNE}),\ }\bibfield  {title}
  {\bibinfo {title} {{Deep Underground Neutrino Experiment (DUNE), Far Detector
  Technical Design Report, Volume I Introduction to DUNE}},\ }\href
  {https://doi.org/10.1088/1748-0221/15/08/T08008} {\bibfield  {journal}
  {\bibinfo  {journal} {JINST}\ }\textbf {\bibinfo {volume} {15}}\bibfield
  {number} {\bibinfo  {number} { (08)},\ \bibinfo {pages} {T08008}},\ }\Eprint
  {https://arxiv.org/abs/2002.02967} {arXiv:2002.02967 [physics.ins-det]}
  \BibitemShut {NoStop}%
\bibitem [{\citenamefont {Abi}\ \emph {et~al.}(2020{\natexlab{b}})\citenamefont
  {Abi} \emph {et~al.}}]{DUNE:2020jqi}%
  \BibitemOpen
  \bibfield  {author} {\bibinfo {author} {\bibfnamefont {B.}~\bibnamefont
  {Abi}} \emph {et~al.} (\bibinfo {collaboration} {DUNE}),\ }\bibfield  {title}
  {\bibinfo {title} {{Long-baseline neutrino oscillation physics potential of
  the DUNE experiment}},\ }\href
  {https://doi.org/10.1140/epjc/s10052-020-08456-z} {\bibfield  {journal}
  {\bibinfo  {journal} {Eur. Phys. J. C}\ }\textbf {\bibinfo {volume} {80}},\
  \bibinfo {pages} {978} (\bibinfo {year} {2020}{\natexlab{b}})},\ \Eprint
  {https://arxiv.org/abs/2006.16043} {arXiv:2006.16043 [hep-ex]} \BibitemShut
  {NoStop}%
\bibitem [{\citenamefont {Alan~Kosteleck\'y}\ and\ \citenamefont
  {Mewes}(2004)}]{PhysRevD.69.016005}%
  \BibitemOpen
  \bibfield  {author} {\bibinfo {author} {\bibfnamefont {V.}~\bibnamefont
  {Alan~Kosteleck\'y}}\ and\ \bibinfo {author} {\bibfnamefont {M.}~\bibnamefont
  {Mewes}},\ }\bibfield  {title} {\bibinfo {title} {Lorentz and cpt violation
  in neutrinos},\ }\href {https://doi.org/10.1103/PhysRevD.69.016005}
  {\bibfield  {journal} {\bibinfo  {journal} {Phys. Rev. D}\ }\textbf {\bibinfo
  {volume} {69}},\ \bibinfo {pages} {016005} (\bibinfo {year}
  {2004})}\BibitemShut {NoStop}%
\bibitem [{\citenamefont {Kosteleck\'y}\ and\ \citenamefont
  {Mewes}(2012)}]{PhysRevD.85.096005}%
  \BibitemOpen
  \bibfield  {author} {\bibinfo {author} {\bibfnamefont {V.~A.}\ \bibnamefont
  {Kosteleck\'y}}\ and\ \bibinfo {author} {\bibfnamefont {M.}~\bibnamefont
  {Mewes}},\ }\bibfield  {title} {\bibinfo {title} {Neutrinos with
  lorentz-violating operators of arbitrary dimension},\ }\href
  {https://doi.org/10.1103/PhysRevD.85.096005} {\bibfield  {journal} {\bibinfo
  {journal} {Phys. Rev. D}\ }\textbf {\bibinfo {volume} {85}},\ \bibinfo
  {pages} {096005} (\bibinfo {year} {2012})}\BibitemShut {NoStop}%
\bibitem [{\citenamefont {Kopp}\ \emph {et~al.}(2008)\citenamefont {Kopp},
  \citenamefont {Lindner}, \citenamefont {Ota},\ and\ \citenamefont
  {Sato}}]{Kopp:2007ne}%
  \BibitemOpen
  \bibfield  {author} {\bibinfo {author} {\bibfnamefont {J.}~\bibnamefont
  {Kopp}}, \bibinfo {author} {\bibfnamefont {M.}~\bibnamefont {Lindner}},
  \bibinfo {author} {\bibfnamefont {T.}~\bibnamefont {Ota}},\ and\ \bibinfo
  {author} {\bibfnamefont {J.}~\bibnamefont {Sato}},\ }\bibfield  {title}
  {\bibinfo {title} {{Non-standard neutrino interactions in reactor and
  superbeam experiments}},\ }\href {https://doi.org/10.1103/PhysRevD.77.013007}
  {\bibfield  {journal} {\bibinfo  {journal} {Phys. Rev. D}\ }\textbf {\bibinfo
  {volume} {77}},\ \bibinfo {pages} {013007} (\bibinfo {year} {2008})},\
  \Eprint {https://arxiv.org/abs/0708.0152} {arXiv:0708.0152 [hep-ph]}
  \BibitemShut {NoStop}%
\bibitem [{\citenamefont {Kosteleck\'y}\ and\ \citenamefont
  {Mewes}(2002)}]{PhysRevD.66.056005}%
  \BibitemOpen
  \bibfield  {author} {\bibinfo {author} {\bibfnamefont {V.~A.}\ \bibnamefont
  {Kosteleck\'y}}\ and\ \bibinfo {author} {\bibfnamefont {M.}~\bibnamefont
  {Mewes}},\ }\bibfield  {title} {\bibinfo {title} {Signals for lorentz
  violation in electrodynamics},\ }\href
  {https://doi.org/10.1103/PhysRevD.66.056005} {\bibfield  {journal} {\bibinfo
  {journal} {Phys. Rev. D}\ }\textbf {\bibinfo {volume} {66}},\ \bibinfo
  {pages} {056005} (\bibinfo {year} {2002})}\BibitemShut {NoStop}%
\bibitem [{\citenamefont {Mishra}\ \emph {et~al.}(2024)\citenamefont {Mishra},
  \citenamefont {Shukla}, \citenamefont {Singh},\ and\ \citenamefont
  {Singh}}]{PhysRevD.109.075042}%
  \BibitemOpen
  \bibfield  {author} {\bibinfo {author} {\bibfnamefont {S.}~\bibnamefont
  {Mishra}}, \bibinfo {author} {\bibfnamefont {S.}~\bibnamefont {Shukla}},
  \bibinfo {author} {\bibfnamefont {L.}~\bibnamefont {Singh}},\ and\ \bibinfo
  {author} {\bibfnamefont {V.}~\bibnamefont {Singh}},\ }\bibfield  {title}
  {\bibinfo {title} {Search for lorentz violations through the sidereal effect
  at the no$\nu$a experiment},\ }\href
  {https://doi.org/10.1103/PhysRevD.109.075042} {\bibfield  {journal} {\bibinfo
   {journal} {Phys. Rev. D}\ }\textbf {\bibinfo {volume} {109}},\ \bibinfo
  {pages} {075042} (\bibinfo {year} {2024})}\BibitemShut {NoStop}%
\bibitem [{\citenamefont {Kosteleck\'y}\ and\ \citenamefont
  {Russell}(2011)}]{RevModPhys.83.11}%
  \BibitemOpen
  \bibfield  {author} {\bibinfo {author} {\bibfnamefont {V.~A.}\ \bibnamefont
  {Kosteleck\'y}}\ and\ \bibinfo {author} {\bibfnamefont {N.}~\bibnamefont
  {Russell}},\ }\bibfield  {title} {\bibinfo {title} {Data tables for lorentz
  and $cpt$ violation},\ }\href {https://doi.org/10.1103/RevModPhys.83.11}
  {\bibfield  {journal} {\bibinfo  {journal} {Rev. Mod. Phys.}\ }\textbf
  {\bibinfo {volume} {83}},\ \bibinfo {pages} {11} (\bibinfo {year}
  {2011})}\BibitemShut {NoStop}%
\bibitem [{\citenamefont {Abi}\ \emph {et~al.}(2021)\citenamefont {Abi} \emph
  {et~al.}}]{DUNE:2021cuw}%
  \BibitemOpen
  \bibfield  {author} {\bibinfo {author} {\bibfnamefont {B.}~\bibnamefont
  {Abi}} \emph {et~al.} (\bibinfo {collaboration} {DUNE}),\ }\bibfield  {title}
  {\bibinfo {title} {{Experiment Simulation Configurations Approximating DUNE
  TDR}},\ }\href@noop {} {\  (\bibinfo {year} {2021})},\ \Eprint
  {https://arxiv.org/abs/2103.04797} {arXiv:2103.04797 [hep-ex]} \BibitemShut
  {NoStop}%
\bibitem [{\citenamefont {Liao}\ \emph {et~al.}(2016)\citenamefont {Liao},
  \citenamefont {Marfatia},\ and\ \citenamefont {Whisnant}}]{Liao:2016hsa}%
  \BibitemOpen
  \bibfield  {author} {\bibinfo {author} {\bibfnamefont {J.}~\bibnamefont
  {Liao}}, \bibinfo {author} {\bibfnamefont {D.}~\bibnamefont {Marfatia}},\
  and\ \bibinfo {author} {\bibfnamefont {K.}~\bibnamefont {Whisnant}},\
  }\bibfield  {title} {\bibinfo {title} {{Degeneracies in long-baseline
  neutrino experiments from nonstandard interactions}},\ }\href
  {https://doi.org/10.1103/PhysRevD.93.093016} {\bibfield  {journal} {\bibinfo
  {journal} {Phys. Rev.}\ }\textbf {\bibinfo {volume} {D93}},\ \bibinfo {pages}
  {093016} (\bibinfo {year} {2016})},\ \Eprint
  {https://arxiv.org/abs/1601.00927} {arXiv:1601.00927 [hep-ph]} \BibitemShut
  {NoStop}%
\bibitem [{\citenamefont {Esteves~Chaves}\ \emph {et~al.}(2018)\citenamefont
  {Esteves~Chaves}, \citenamefont {Rossi~Gratieri},\ and\ \citenamefont
  {Peres}}]{Chaves:2018sih}%
  \BibitemOpen
  \bibfield  {author} {\bibinfo {author} {\bibfnamefont {M.}~\bibnamefont
  {Esteves~Chaves}}, \bibinfo {author} {\bibfnamefont {D.}~\bibnamefont
  {Rossi~Gratieri}},\ and\ \bibinfo {author} {\bibfnamefont {O.~L.~G.}\
  \bibnamefont {Peres}},\ }\bibfield  {title} {\bibinfo {title} {{Improvements
  on perturbative oscillation formulas including non-standard neutrino
  Interactions}},\ }\href@noop {} {\  (\bibinfo {year} {2018})},\ \Eprint
  {https://arxiv.org/abs/1810.04979} {arXiv:1810.04979 [hep-ph]} \BibitemShut
  {NoStop}%
\bibitem [{\citenamefont {Yasuda}(2007)}]{Yasuda:2007jp}%
  \BibitemOpen
  \bibfield  {author} {\bibinfo {author} {\bibfnamefont {O.}~\bibnamefont
  {Yasuda}},\ }\bibfield  {title} {\bibinfo {title} {{On the exact formula for
  neutrino oscillation probability by Kimura, Takamura and Yokomakura}},\
  }\href@noop {} {\  (\bibinfo {year} {2007})},\ \Eprint
  {https://arxiv.org/abs/0704.1531} {arXiv:0704.1531 [hep-ph]} \BibitemShut
  {NoStop}%
\bibitem [{\citenamefont {Baker}\ and\ \citenamefont
  {Cousins}(1984)}]{Baker:1983tu}%
  \BibitemOpen
  \bibfield  {author} {\bibinfo {author} {\bibfnamefont {S.}~\bibnamefont
  {Baker}}\ and\ \bibinfo {author} {\bibfnamefont {R.~D.}\ \bibnamefont
  {Cousins}},\ }\bibfield  {title} {\bibinfo {title} {{Clarification of the Use
  of Chi Square and Likelihood Functions in Fits to Histograms}},\ }\href
  {https://doi.org/10.1016/0167-5087(84)90016-4} {\bibfield  {journal}
  {\bibinfo  {journal} {Nucl. Instrum. Meth.}\ }\textbf {\bibinfo {volume}
  {221}},\ \bibinfo {pages} {437} (\bibinfo {year} {1984})}\BibitemShut
  {NoStop}%
\end{thebibliography}%
\end{document}